\documentclass[aps,prb,reprint,superscriptaddress,twocolumn]{revtex4-2}

\usepackage{bm,mathrsfs,dcolumn,graphicx,color} 
\usepackage{amsmath}
\usepackage{graphicx}
\usepackage{tabularx}
\usepackage[T1]{fontenc}
\usepackage[english]{babel}
\usepackage{dsfont}
\usepackage{xcolor}
\usepackage[utf8]{inputenc}
\usepackage{amssymb}
\usepackage{braket}

\usepackage[colorlinks=true, allcolors=blue]{hyperref}

\renewcommand{\bra}[1]{\left< #1 \right|}
\renewcommand{\ket}[1]{\left| #1 \right>}

\newcommand{\bq}{\mathbf{q}}

\begin{document}

\title{Impact of phonon lifetimes on the single-photon indistinguishability in quantum emitters based on 2D materials}

\author{Alexander Steinhoff}
\email{asteinhoff@itp.uni-bremen.de}
\affiliation{Institut für Theoretische Physik, Universität Bremen, 28334 Bremen, Germany}
\affiliation{Bremen Center for Computational Materials Science, Universit\"at Bremen, 28334 Bremen, Germany}
\author{Steffen Wilksen}
\affiliation{Institute for Physics, Carl von Ossietzky Universität Oldenburg}
\author{Ivan Solovev}
\affiliation{Institute for Physics, Carl von Ossietzky Universität Oldenburg}
\author{Christian Schneider}
\affiliation{Institute for Physics, Carl von Ossietzky Universität Oldenburg}
\author{Christopher Gies}
\affiliation{Institute for Physics, Carl von Ossietzky Universität Oldenburg}

\begin{abstract}

Localized excitons in two-dimensional (2D) materials are considered as promising sources of single photons on demand. The photon indistinguishability as key figure of merit for quantum information processing is strongly influenced by the coupling of charge excitations to lattice vibrations of the surrounding semiconductor material. Here,  we quantify the impact of exciton-acoustic-phonon-interaction and cavity QED effects on photon indistinguishability in a Hong-Ou-Mandel setup by solving fully quantum mechanical equations for the coupled QD-cavity-phonon system including non-Markovian effects. We find a strong reduction of indistinguishability compared to 3D systems due to increased exciton-phonon coupling efficiency. Moreover, we show that the coherence properties of photons are significantly influenced by the finite phonon lifetime in the surrounding material giving rise to pure dephasing. Only if these limitations are overcome, localized excitons in 2D semiconductors can become a new avenue for quantum light sources.
\end{abstract}

\maketitle

\section{Introduction}

Single photon sources are an essential building block for quantum technologies, enabling a variety of applications such as quantum information transfer \cite{kimble_quantum_2008, aharonovich_solid-state_2016, gao_atomically-thin_2023}, entanglement generation \cite{flissikowski_two-photon_2004, jayakumar_deterministic_2013}, random number generation \cite{ma_quantum_2016}, photonic computing \cite{obrien_photonic_2009, wang_boson_2019}, and much more. While the sought after quality for quantum key distribution and random number generation is the purity of the single-photon emission as quantified by the antibunching, i.e.\ the diminishing likelihood of more than a single photon being emitted at each instance in time, a majority of the advanced quantum-technology applications relies on quantum interference. Especially in photonic computing, gate fidelity essentially relies on the precise spatio-temporal overlap of photon wave packets at the linear optical elements \cite{lal_indistinguishable_2022, wang_boson_2019, peyskens_integration_2019}. The capability of quantum objects to exhibit quantum interference effects relies on their indistinguishability, which is quantified in terms of a self-interference effect at a balanced beam splitter -- the well know Hong-Ou-Mandel measurement \cite{hong_measurement_1987, santori_indistinguishable_2002, wang_towards_2019}. 

To this day, the reign of on-demand, bright solid state sources of indistinguishable photons is held by epitaxially grown semiconductor nanostructures, such as III/V quantum dots (QDs), and other platform such as silicon defects \cite{komza_indistinguishable_2024}.
%and ... [Alex, baust Du die vielleicht noch ein: Zhang et al., Sci. Adv. 8, eabn9252 (2022), Gazzano DOI: 10.1038/ncomms2434 zu QDs aus Senellart Gruppe]. 
In these systems, single-photon purity and indistinguishability above 99\% have been achieved \cite{wei_deterministic_2014}. In recent years, two-dimensional (2D) semiconductor materials have been shown to serve well as cost-effective single-photon sources \cite{yu_tunable_2024}. Instead of advanced nanoprocessing and epitaxy, quantum-dot like emission properties can be achieved via localized strain, such as resulting from nano- and microstructures \cite{kern_nanoscale_2016, palacios-berraquero_large-scale_2017, yu_site-controlled_2021, wang_highly_2021, paralikis_tailoring_2024} or nanoparticles \cite{tripathi_spontaneous_2018}, on which the atomically thin monolayer is placed, or nanobubbles \cite{shepard_nanobubble_2017, cianci_spatially_2023}, where single-photon emission can arise due to various effects \cite{darlington_imaging_2020}. Also defects in hexagonal boron nitride have been demonstrated to give rise to localized states required for single-photon emission \cite{preuss_resonant_2022, aharonovich_quantum_2022, samaner_freespace_2022, kretzschmar_quantitative_2024}. While relatively high single-photon purity has been demonstrated in these materials, indistinguishability is rarely being addressed \cite{drawer_monolayer-based_2023,fournier_two-photon_2023}.

%, possibly for the reason that it is lacking far behind its III/V competitors \cite{}[Nanophotonics 2024; 13(19): 3615–3629].

Only recently, it was brought to the attention of the community working on 2D single photon sources that the interaction with phonons 
%in the 2D layered material systems 
may pose a fundamental limitation to attainable photon indistinguishability that might not be straightforward to overcome by simple improvements in fabrication \cite{vannucci_single-photon_2024, mitryakhin_engineering_2024}.
%
%While 
%quasi-2D quantum-well structures have been around for decades \TODO{should we put a Ref here? Ja!}, the situation is different for van der Waals (vdW) materials such as transition metal dichalcogenide (TMD) semiconductors. The embedding of quantum wells into the surrounding bulk material effectively yields a coupling of the active material to phonons in three dimensions. On the other hand, 
%
While III/V QD emitters couple to 3D phonons of the surrounding bulk material,
in vdW monolayers phonons truly exhibit a 2D density of states (DOS), fundamentally  impacting the emitter-phonon coupling efficiency. 
%A similar prediction has been made for emitters in a one-dimensional environment \cite{ferreira_neto_one-dimensional_2024}. 
%, strongly enhancing their detrimental impact on the single-photon emission quality due to increased coupling efficiency at long wavelengths.
Various signatures of phonons in 2D materials have recently been
investigated \cite{khatri_phonon_2019,denning_quantum_2022,preuss_resonant_2022,svendsen_signatures_2023,mitryakhin_engineering_2024,piccinini_high-purity_2024,vannucci_single-photon_2024}. 
%while a one-dimensional (1D) phonon spectral density has been utilized in \cite{kikas_anomalous_1996,reichman_nonperturbative_1996,liu_dissipation_2020,laferriere_approaching_2023, ferreira_neto_one-dimensional_2024}.
In particular, the photon indistinguishability  $\mathcal{I}$ is very sensitive to phonon-induced dephasing processes. In a simple Markovian picture, one immediately recognizes the effect of pure dephasing on $\mathcal{I}$ \cite{grange_cavity-funneled_2015}. Pure dephasing yields a broadening of the emitter zero phonon line (ZPL) and is related to the exponential long-time decay of coherences, caused by phonons with small momenta \cite{krummheuer_theory_2002}. While in 1D, linear coupling to longitudinal acoustic (LA) phonons gives rise to pure dephasing \cite{ferreira_neto_one-dimensional_2024}, this is not the case in 2D systems.
%quadratic coupling in 2d to LA phonons yields pure dephasing \cite{vannucci_single-photon_2024}
However, even in the absence of pure dephasing, phonon-induced dephasing is much stronger in two than in three dimensions due to non-Markovian phonon side band (PSB) effects
\cite{kaer_microscopic_2013, mitryakhin_engineering_2024} impacting the short-time behavior of coherences.

In this paper, we quantify the single-photon indistinguishability $\mathcal{I}$ by modeling a HOM setup of a QD embedded in monolayer WSe$_2$ coupled to a cavity and subject to the interaction with 2D LA phonons. To this end, we use emitter-phonon coupling parameters obtained from experiment \cite{mitryakhin_engineering_2024} as input to a master-equation approach by solving the von Neumann-Lindblad (vNL) equation. Calculations are performed for a temperature $T\rightarrow 0$ K and zero cavity-emitter detuning. The phonons are included on a fully quantum-mechanical footing as pioneered by Hohenester \cite{hohenester_quantum_2007} and Kaer et al. \cite{kaer_microscopic_2013}, thereby taking into account non-Markovian effects as well as multi-phonon processes. We include phonon lifetime effects by means of additional Lindblad terms describing phonon decay. As revealed by the solution of a modified independent boson model (IBM) \cite{zimmermann_dephasing_2002}, phonon decay turns out to be a natural source of pure dephasing. We therefore quantify the impact of the phonon lifetime as an external parameter imposed by the experimental setup. Besides by the pure dephasing rate, $\mathcal{I}$ can be influenced by tuning the radiative emitter lifetime via engineering of the cavity quality factor $Q$ as well as the photon DOS, which is approximately represented by the Purcell factor $F_{\textrm{P}}$. To assess whether cavities can be used to overcome the limitations to the indistinguishability one can expect of truly 2D QEs, we study the dependence of $\mathcal{I}$ on the cavity parameters for a wide range of $Q$ and $F_{\textrm{P}}$. Our spectrum of model parameters lies well within the range of values that are achievable in current experiments on quantum emitters in 2D systems, some of which we have collected in Table~\ref{table:cavities}, but also reach out to the regime of improved setups in near-future realizations.
We find that although the detrimental impact of dephasing may be mitigated by choosing an appropriate cavity, a fundamental limitation to values $\mathcal{I}\approx 0.3-0.4$ remains. 
%\TODO{we might use parts the following two paragraphs, which are condensed from the respective references:}
%\\\cite{vannucci_single-photon_2024}: predictions on the excitation dynamics of
%the system under different schemes — resonant excitation, phonon-assisted pumping, and SUPER swingup; results show the signature of strong exciton-phonon coupling
%in WSe2; also: fundamental limitations dictated by phonon scattering to the single-photon indistinguishability by analyzing the phonon-induced broadening of the ZPL within a simple pure-dephasing picture assuming filtering of the PSB.
%\\
%\\\cite{ferreira_neto_one-dimensional_2024}: phonon-induced decoherence for a quantum dot placed in the one-dimensional (1D) system of a homogeneous cylindrical nanowire. 
%consider both a linear and a quadratic coupling of the emitter to these modes. analytical expression for the 1D pure dephasing rate, which leads to a reduced pure dephasing rate compared with bulk. By implementing these results into a full cavity quantum electrodynamic model, we demonstrate that multimode coupling is necessary to
%correctly predict the indistinguishability in a 1D system, which may otherwise be significantly underestimated; importance of going beyond the assumption of Markovian decay in a 1D photonic wire, where memory and back-action effects in the reservoir can increase photon indistinguishability.
%
\begin{table}
\begin{tabularx}{8cm}{|X|c|c|}
\hline
\textrm{Architecture} & \textrm{Quality factor} & \textrm{Purcell factor} \\ 
\hline
% monolithic microcavity 
% & 8000 \cite{Yang2024} 
% & 9 \cite{Yang2024} 
% & \cite{Yang2024} 
% \\ 
% \hline
Open microcavity & & \\  
WSe$_2$ \cite{Flatten2018}  & 545& 18 
\\  
WSe$_2$ \cite{drawer_monolayer-based_2023} &  600  & 1.3 
\\    \hline
Circular Bragg grating  & & \\   
WSe$_2$  \cite{Iff2021} & 40 & 5 
\\      \hline
Plasmonic structure  & & \\   
hexagonal boron nitride \cite{Tran2017}  & 12 
& 31$^*$ \\ 
hexagonal boron nitride \cite{Sakib2024} & 8$^{\dagger}$ & 10 \\
WSe$_2$ \cite{Luo2018} & 8 & 181
\\
\hline
\end{tabularx}
\label{table:cavities}
\caption{The experimentally measured performance of quantum emitters in 2D systems embedded in various cavity platforms. * Upper limit. $\dagger$ Our estimate based on the emission profile.}
\end{table}
\begin{figure}
\centering
\includegraphics[width=\columnwidth]{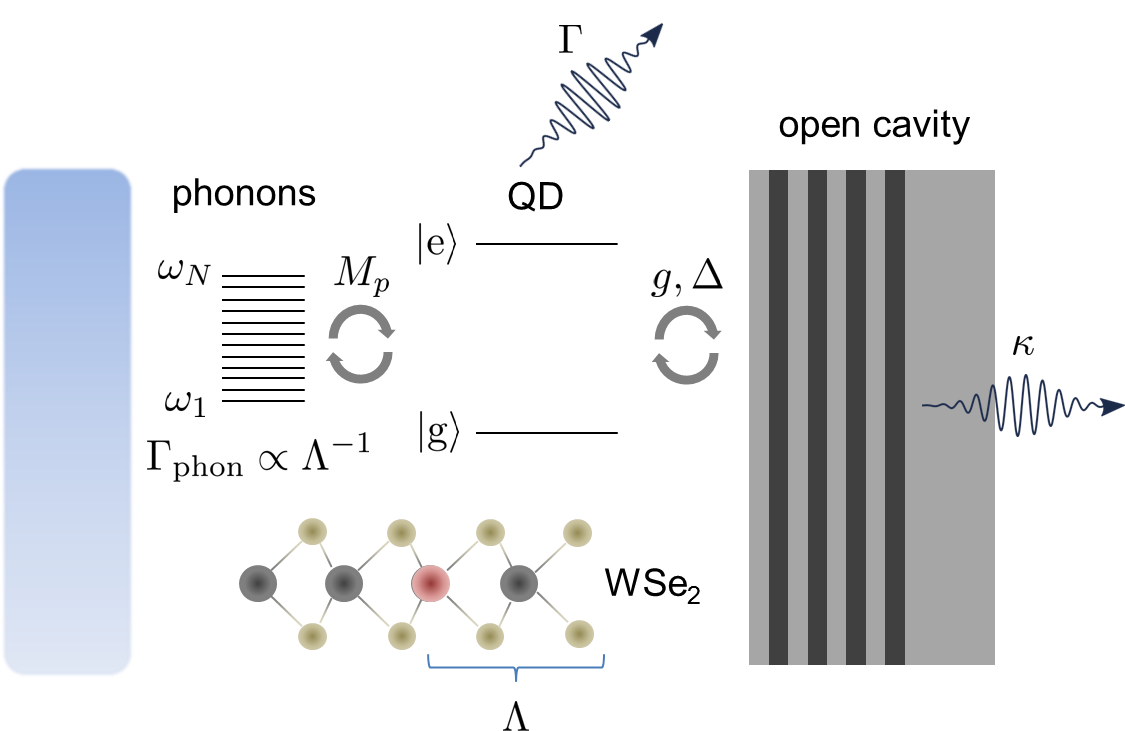}
\caption{Schematic of a WSe$_2$ QD coupled to $N$ 2D longitudinal acoustic phonon modes via matrix elements $M_q$ and to an exemplary open cavity via the Jaynes-Cummings parameter $g$. The detuning between the QD transition and the cavity photon mode is given by $\Delta=\omega_{\textrm{e}}-\omega_{\textrm{g}}-\omega_{\textrm{cav}}$. Spontaneous emission of the QD into other modes is described by the rate $\Gamma$ and cavity losses take place at the rate $\kappa$. The phonons decay at a rate $\Gamma_{\textrm{phon}}$ inversely proportional to the phonon mean free path $\Lambda$, which is restricted to the smallest distance between the QD (red dot) and sample boundaries.}
\label{fig:model}
\end{figure}

\section{Theory}

\begin{figure}
\centering
\includegraphics[width=1.\columnwidth]{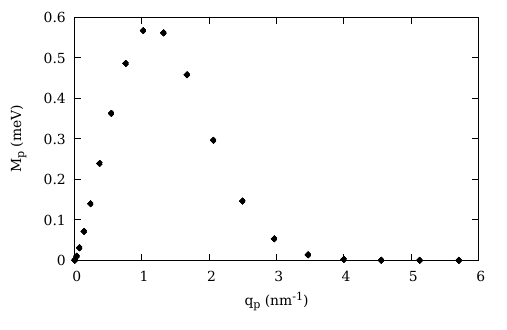}
\caption{Coupling matrix elements $M_p$ between WSe$_2$ QD excitons and 2D LA phonons depending on the momentum $q_p$ of the effective discrete phonon modes.}
\label{fig:M_k}
\end{figure}

Our model of a quantum dot (QD) embedded in monolayer WSe$_2$ and coupled to 2D longitudinal acoustic (LA) phonons as well as the photon mode of a cavity is schematically shown in Fig.~\ref{fig:model}. The QD-cavity system is described by a Jaynes-Cummings Hamiltonian expanded in the basis $\left\{ \ket{1}=\ket{\textrm{e},n=0}, \ket{2}=\ket{\textrm{g},n=1},\ket{3}=\ket{\textrm{g},n=0} \right\}$, where $n$ is the cavity photon number:
\begin{equation}
 \begin{split}
H_{\textrm{JC}}&=\hbar\Delta\sigma_{11}+\hbar g(\sigma_{12}+\sigma_{21}) \,.
\end{split}
\label{eq:Hamiltonian_JC}
\end{equation}
Here, $\sigma_{ij}=\ket{i}\bra{j}$ and the detuning between QD and cavity is given by $\Delta=\omega_{\textrm{eg}}-\omega_{\textrm{cav}}$ with the transition frequencies $\omega_{\textrm{eg}}$ and $\omega_{\textrm{cav}}$ of the QD and the cavity, respectively. Following Refs.~\cite{hohenester_quantum_2007, kaer_microscopic_2013}, we introduce a set of effective discrete phonon modes $\left\{\ket{p}\right\}$, to which the excited QD state couples via matrix elements $M_p$. The total Hamiltonian is given by
\begin{equation}
 \begin{split}
H=H_{\textrm{JC}}+\sum_p M_p (b^{\dagger}_p+b^{\phantom\dagger}_p)\sigma_{11}+\sum_p \hbar\omega_p b^{\dagger}_p b^{\phantom\dagger}_p\,,
\end{split}
\label{eq:Hamiltonian_tot}
\end{equation}
where $b^{\dagger}_p\, (b^{\phantom\dagger}_p)$ creates (annihilates) a phonon in mode $\ket{p}$. We assume that exciton-phonon coupling in TMD monolayers is dominated at low temperatures by longitudinal acoustic (LA) phonons with linear dispersion $\omega_p=c_{\textrm{ac}} q_p$, where $q_p$ is the absolute value of phonon momentum for mode $p$. The effective matrix elements for QD excitons coupling to 2D LA phonons are given by 
\begin{equation}
 \begin{split}
M_p=\Big(2\pi\Delta_p q_p \frac{\mathcal{A}}{(2\pi)^2}\Big)^{1/2} \hbar g_{\textrm{ac},q_p}\,,
\end{split}
\label{eq:X_phon_effective_ME}
\end{equation}
where the microscopic matrix elements are
\begin{equation}
 \begin{split}
g_{\textrm{ac},q}= \Big(\frac{q}{2\rho \hbar c_{\textrm{ac}} \mathcal{A}}\Big)^{1/2}(D_{\textrm{e}} - D_{\textrm{h}})e^{-\frac{1}{4}(ql_{\textrm{QD}})^2} \,.
\end{split}
\label{eq:X_phon_microscopic_ME}
\end{equation}
Here, $\rho$ is the 2D mass density, $l_{\textrm{QD}}$ is the extension of the electron and hole wavefunctions, $D_{\textrm{e/h}}$ are the electron and hole deformation potentials, respectively, and $\mathcal{A}$ is the crystal area. 
As discussed in detail in Ref.~\cite{mitryakhin_engineering_2024}, numerical values for the parameters are obtained by consistently fitting a model for the coupled emitter-phonon system to a series of experimental spectra. Parameters are collected in Table~\ref{table:Xphon}. 
For simplicity, we use $l_{\textrm{QD}}=(l_e+l_h)/2$. The phonon momenta are discretized on a Chebyshev-Gauss-type grid $q_p,\, p=1,...,N$ (see Appendix \ref{app:num_details} for details), which is dense around $q=0$ and becomes gradually coarser for increasing momenta. In Eq.~(\ref{eq:X_phon_effective_ME}), the weighting factors $\Delta_p$ for the effective phonon modes are given by $\Delta_p=q_{p+1}-q_p$. The resulting effective matrix elements are shown in Fig.~\ref{fig:M_k}.
As many-particle basis for the phonons, we use the occupation number states $\ket{\left\{ n_p \right\}}=\ket{n_1,n_2,...,n_N} $ with
\begin{equation}
 \begin{split}
 b^{\dagger}_{p'} \ket{\left\{ n_p \right\}} & = \sqrt{n_{p'}+1} \ket{\left\{ n_p \right\}, n_{p'}+1}\,, \\
 b_{p'} \ket{\left\{ n_p \right\}} & = \sqrt{n_{p'}} \ket{\left\{ n_p \right\}, n_{p'}-1}
 \,.
\end{split}
\label{eq:occ_number_states}
\end{equation}
The basis for the coupled QD-cavity-phonon system is given by product states $\ket{i}\otimes \ket{\left\{ n_p \right\}}=\ket{i,\left\{ n_p \right\}}$, which exist in a Fock space of variable total phonon number. Full quantum information of the coupled system is contained in the density matrix $\hat{\rho}=\sum_{i,j,p,p'}\ket{i,\left\{ n_p \right\}} \hat{\rho}^{i,j}_{p,p'} \bra{j,\left\{ n_{p'} \right\}} $.
Any expectation value for the Jaynes-Cummings (JC) subspace can be obtained by a partial trace over the phonon degrees of freedom:
\begin{equation}
 \begin{split}
\left< A_{\textrm{JC}} \right>=\textrm{tr}_{\textrm{JC}}\textrm{tr}_{\textrm{ph}}\left\{ A_{\textrm{JC}}\mathds{1}_{\textrm{ph}} \hat{\rho}  \right\}=
\textrm{tr}_{\textrm{JC}}\left\{ A_{\textrm{JC}}\hat{\rho}_{\textrm{JC}} \right\}\,,
\end{split}
\label{eq:partial_trace}
\end{equation}
where $\hat{\rho}_{\textrm{JC}}=\sum_{i,j}\ket{i} \sum_{p}\hat{\rho}^{i,j}_{p,p} \bra{j} $ is the reduced density matrix of the QD-cavity system.
\begin{table}
\begin{tabularx}{5cm}{|X|r|}
\hline
$D_{\textrm{e}}$ (eV) &                        $3.82$ \\ 
\hline
$D_{\textrm{h}}$ (eV) &                        $1.07$ \\    
\hline
$c_{\textrm{ac}}$ (nm ps$^{-1}$) &   $3.92$ \\     
\hline
$l_e$ (nm) &                         $1.36$  \\  
\hline
$l_h$ (nm) &                         $1.42$ \\ 
\hline
$\rho$ (meV ps$^2$ nm$^{-4}$) &      $3.69\times10^4$     \\   
\hline
\end{tabularx}
\label{table:Xphon}
\caption{Parameters for the exciton-phonon coupling taken from \cite{mitryakhin_engineering_2024}.}
\end{table}

The dynamics of the system is governed by the von Neumann-Lindblad (vNL) equation for the density matrix $\hat{\rho}$, where dissipation due to coupling of the QD-cavity-phonon system to the environment is described by Lindblad terms \cite{lindblad_generators_1976,breuer_theory_2007}:
\begin{equation}
 \begin{split}
\frac{\partial}{\partial t}\hat{\rho}(t)= -\frac{i}{\hbar}\left[H, \hat{\rho}(t) \right]+\mathcal{L}\hat{\rho}(t)\,.
\end{split}
\label{eq:vNL}
\end{equation}
The Lindblad terms are given by $\mathcal{L}\hat{\rho}(t)=\big[\mathcal{L}_{\kappa}(\sigma_{32})+\mathcal{L}_{\Gamma}(\sigma_{31})+\sum_p \mathcal{L}_{\Gamma_{\textrm{phon}}}(b_p)\big]\hat{\rho}(t) $
with the general form
$\mathcal{L}_{\gamma}(\hat{O})\hat{\rho}=-\frac{\gamma}{2}\big[
\hat{O}^{\dagger}\hat{O}\hat{\rho}+\hat{\rho}\hat{O}^{\dagger}\hat{O}-2\hat{O}\hat{\rho}\hat{O}^{\dagger}
\big] $.
The rate $\Gamma$ describes the decay of QD excitations via spontaneous emission of photons in the absence of the cavity, $\kappa$ is the rate of photon decay from the cavity, and $\Gamma_{\textrm{phon}}$ is the decay rate of phonons.
From Eq.~(\ref{eq:vNL}), we derive coupled EOM for the individual matrix elements of $\hat{\rho}$ via $\hat{\rho}^{i,j}_{p,p'}=\bra{i,\left\{ n_{p} \right\}}\hat{\rho}\ket{j,\left\{ n_p' \right\}}$, which are solved numerically. To this end, we truncate the Fock space by allowing for a maximum total phonon number $N_{\textrm{ph}}$. 
We discuss the convergence of numerical results with respect to the many-particle phonon basis in the Appendix.
For all calculations, we assume that no phonons are excited in the initial state, corresponding to a lattice temperature of $T=0$ K.

For the WSe$_2$ QD emitter, we use parameters from Ref.~\cite{mitryakhin_engineering_2024}: The zero-phonon-line (ZPL) energy is given by $1.6$ eV, hence $\omega_{\textrm{cav}}=2.4\times 10^3$ ps$^{-1}$ for a cavity that is in resonance with the emitter. The radiative emitter lifetime including Purcell enhancement from the cavity is $\tau_{\textrm{cav}}=1.89$ ns. From the characterization of a similar cavity \cite{drawer_monolayer-based_2023}, we estimate a Purcell factor of $F_{\textrm{P}}=1.5$. Therefore, the radiative emitter lifetime in free space is given by
\begin{equation}\label{eq:tau_free}
\tau_{\textrm{free}}=\tau_{\textrm{cav}} F_{\textrm{P}}=\Gamma^{-1}
\end{equation}
and accounting to a value of 2.8~ns. In the following, we investigate cavities with varying quality (Q-)factors. The Q-factor is defined as the ratio between cavity frequency and (FWHM) linewidth, where the latter is given by the cavity decay rate $\kappa$ entering the vNL equation:
\begin{equation}
 \begin{split}
Q=\frac{\omega_{\textrm{cav}}}{\kappa}\,.
\end{split}
\label{eq:Q}
\end{equation}
A model for the phonon lifetime has been introduced in Ref.~\cite{zimmermann_dephasing_2002}. While at finite temperatures, phonon-phonon scattering yields a characteristic temperature dependence of the phonon lifetime, all scattering channels are frozen at $T=0$ K. In this case, phonons only decay due to a limited phonon free mean path $\Lambda$, which is restricted to the smallest distance between the QD and the sample surface or grain boundaries. A lower bound for the phonon lifetime is then given by
\begin{equation}
 \begin{split}
\Gamma_{\textrm{phon}} = \tau_{\textrm{phon}}^{-1}=\frac{c_{\textrm{ac}}}{\Lambda}\,,
\end{split}
\label{eq:Gamma_phon}
\end{equation}
which is taken into account for each phonon mode in Eq.~(\ref{eq:vNL}) within the Lindblad formalism.

In a Hong-Ou-Mandel (HOM) setup, the indistinguishability $\mathcal{I}$ of photons emitted from the QD is quantified as the normalized number of coincidence events at the detectors. It can be calculated from expectation values of the QD decay operator $\sigma_-=\sigma_{31}$ as \cite{kaer_microscopic_2013}:
\begin{equation}
 \begin{split}
\mathcal{I}=\frac{\int_0^{\infty}\,dt\int_0^{\infty}\,d\tau\,|\big< \sigma_-^{\dagger}(t+\tau)\sigma_-(t) \big>|^2}{\int_0^{\infty}\,dt\int_0^{\infty}\,d\tau\,\big< \sigma_-^{\dagger}(t+\tau)\sigma_-(t+\tau) \big>\big< \sigma_-^{\dagger}(t)\sigma_-(t) \big>}\,.
\end{split}
\label{eq:I}
\end{equation}
The occurring two-time expectation values can be evaluated with the help of the quantum regression formula (QRF) \cite{carmichael_statistical_1999}: 
%%
%\begin{equation}
% \begin{split}
%&\big< A_{\textrm{JC}}(t+\tau)B_{\textrm{JC}}(t) \big> \\
%&=\textrm{tr}\left\{ A_{\textrm{JC}}(t+\tau)B_{\textrm{JC}}(t)\mathds{1}_{\textrm{ph}} \hat{\rho}  \right\} \\
%&=\textrm{tr}\left\{U^{\dagger}(t+\tau) A_{\textrm{JC}}U(t+\tau)U^{\dagger}(t)B_{\textrm{JC}}U(t)\mathds{1}_{\textrm{ph}} \hat{\rho}(0) \right\} \\
%&=\textrm{tr}\left\{A_{\textrm{JC}}U(\tau)B_{\textrm{JC}}\mathds{1}_{\textrm{ph}} \hat{\rho}(t)U^{\dagger}(\tau) \right\} \\
%&=\textrm{tr}_{\textrm{JC}}\left\{A_{\textrm{JC}}\textrm{tr}_{\textrm{ph}}\left\{U(\tau)B_{\textrm{JC}}\mathds{1}_{\textrm{ph}} \hat{\rho}(t)U^{\dagger}(\tau)\right\} \right\} \\
%&=\textrm{tr}_{\textrm{JC}}\left\{A_{\textrm{JC}} \textrm{tr}_{\textrm{ph}}\left\{\hat{\tilde{\rho}}_t(\tau)\right\} \right\}
%\,,
%\end{split}
%\label{eq:QRF}
%\end{equation}
%%
%where we have used the group property of the time evolution operator $U(t)$ and the cyclic property of the trace. 
%
\begin{equation}
 \begin{split}
&\big< A_{\textrm{JC}}(t+\tau)B_{\textrm{JC}}(t) \big> 
=\textrm{tr}_{\textrm{JC}}\left\{A_{\textrm{JC}} \textrm{tr}_{\textrm{ph}}\left\{\hat{\tilde{\rho}}_{B;t}(\tau)\right\} \right\}
\,.
\end{split}
\label{eq:QRF}
\end{equation}
The new density matrix $\hat{\tilde{\rho}}_{B;t}(\tau)$ results from the time evolution in $\tau$ according to a vNL equation similar to Eq.~(\ref{eq:vNL}), starting from the initial matrix $\hat{\tilde{\rho}}_{B;t}(0)=B_{\textrm{JC}}\mathds{1}_{\textrm{ph}} \hat{\rho}(t) $.
For a given operator $B_{\textrm{JC}}=\sigma_{kl}$, the matrix elements of the latter can be related to the original density matrix at time $t$:
\begin{equation}
 \begin{split}
\hat{\tilde{\rho}}^{i,j}_{\sigma_{kl};t;p,p'}(0)&=\bra{i,\left\{ n_{p} \right\}}B_{\textrm{JC}}\mathds{1}_{\textrm{ph}} \hat{\rho}(t)\ket{j,\left\{ n_p' \right\}} \\
&=\bra{i,\left\{ n_{p} \right\}}\mathds{1}_{\textrm{ph}}\ket{k}\bra{l} \hat{\rho}(t)\ket{j,\left\{ n_p' \right\}} \\
&=\delta_{i,k}\hat{\rho}^{l,j}_{p,p'}(t)
\,.
\end{split}
\label{eq:rho_tilde_0}
\end{equation}
With $A_{\textrm{JC}}=B^{\dagger}_{\textrm{JC}}=\sigma_{lk}$, we find from Eq.~(\ref{eq:QRF}):
\begin{equation}
 \begin{split}
&\big< \sigma_{kl}^{\dagger}(t+\tau)\sigma^{\phantom\dagger}_{kl}(t) \big> \\
&=\textrm{tr}_{\textrm{JC}}\left\{\ket{l}\bra{k} \textrm{tr}_{\textrm{ph}}\left\{\hat{\tilde{\rho}}_{\sigma_{kl};t}(\tau)\right\} \right\} \\
&=\sum_p \hat{\tilde{\rho}}^{k,l}_{\sigma_{kl};t;p,p}(\tau) \\
&=\hat{\tilde{\rho}}^{k,l}_{\sigma_{kl};t;\textrm{JC}}(\tau)
\,.
\end{split}
\label{eq:QRF_special}
\end{equation}
Hence, the calculation of $\mathcal{I}$ requires a two-step time evolution in $t$ and $\tau$.
As initial state, we assume the QD to be in the excited state in the absence of photons.

%For the sake of an analytic expression for the polaron shift, as well as to facilitate a reference calculation to our results in Section~\ref{sec:results}, 
To quantify the phonon-induced polaron shift of the ZPL and thereby determine the optimal spectral position of the cavity in our calculations of $\mathcal{I}$, 
it is useful to characterize the coupled QD-phonon system without cavity. We start by deriving the linear optical absorption spectrum as the Fourier transform of the normalized dipole autocorrelation \cite{chang_non-markovian_1993}:
\begin{equation}
 \begin{split}
\alpha(\omega)&=\frac{1}{\left< \mu^2 \right>}\textrm{Re}\,\int_0^{\infty} d\tau e^{i\omega\tau} \left< \mu(\tau)\mu(0)\right>
\end{split}
\label{eq:spectrum1}
\end{equation}
with the dipole operator $\mu = d \sigma_{13} + d^* \sigma_{31}$. Making use of the QRF (\ref{eq:QRF_special}), we find
\begin{equation}
 \begin{split}
\alpha(\omega)=\textrm{Re}\,\int_0^{\infty} d\tau e^{i\omega\tau}\Big[ & 
\hat{\tilde{\rho}}^{1,3}_{\sigma_{13};0;\textrm{JC}}(\tau)
+\hat{\tilde{\rho}}^{1,3}_{\sigma_{31};0;\textrm{JC}}(\tau)
\\
+&\hat{\tilde{\rho}}^{3,1}_{\sigma_{31};0;\textrm{JC}}(\tau)
+\hat{\tilde{\rho}}^{3,1}_{\sigma_{13};0;\textrm{JC}}(\tau)
\Big]
\,.
\end{split}
\label{eq:spectrum2}
\end{equation}
In this case, we use the unexcited QD as initial state.
%
%In analogy to the cavity spectrum \cite{del_valle_luminescence_2009}, which describes the creation of cavity photons, we derive the absorption as the spectral distribution of QD exciton creation at a certain time $t$:
%%
%\begin{equation}
% \begin{split}
%\alpha(\omega;t)&=\frac{1}{\left< \sigma^{\dagger}_{31}(t)\sigma^{\phantom\dagger}_{31}(t) \right>}\textrm{Re}\,\int_0^{\infty} d\tau e^{i\omega\tau} \left< \sigma^{\dagger}_{31}(t)\sigma^{\phantom\dagger}_{31}(t+\tau)\right> \\
%&=\frac{1}{\hat{\rho}_{\textrm{JC}}^{1,1}(t)}\textrm{Re}\,\int_0^{\infty} d\tau e^{i\omega\tau} (\hat{\tilde{\rho}}^{3,1}_{t;\textrm{JC}}(\tau))^*
%\,,
%\end{split}
%\label{eq:spectrum}
%\end{equation}
%%
%where we made use of the QRF (\ref{eq:QRF_special}).

Absorption spectra can be compared to the exactly solvable independent boson model (IBM) augmented by phonon damping, cf.\ Ref.~\cite{zimmermann_dephasing_2002}. This augmented IBM yields explicit expressions for the polaron shift
\begin{equation}
 \begin{split}
\Delta_{\textrm{pol}}&=-\sum_{\bq} g^2_{\textrm{ac},{\bq}} \frac{\omega_{\bq}}{\omega_{\bq}^2+(\gamma_{\bq}/2)^2} \\ &=-\sum_{p}\frac{M^2_{p}}{\hbar^2}\frac{\omega_p}{\omega_p^2+(\Gamma_{\textrm{phon}}/2)^2}\,,
\end{split}
\label{eq:polaron_shift}
\end{equation}
where the mode-dependent damping rate $\gamma_{\bq}$ has been chosen as a constant given by the Lindblad rate $\Gamma_{\textrm{phon}}$, and the afore-mentioned effective phonon modes have been used. 
As shown in Appendix~\ref{app:IBM}, \textit{half} the Lindblad rate enters as characteristic decay rate.
The polaron shift is accompanied by a pure dephasing rate (at zero temperature)
\begin{equation}
 \begin{split}
\frac{\gamma_{\textrm{pure}}}{2}&=\sum_{\bq}g^2_{\textrm{ac},{\bq}}\frac{\gamma_{\bq}/2}{\omega_{\bq}^2+(\gamma_{\bq}/2)^2} \\
&=\sum_{p}\frac{M^2_{p}}{\hbar^2}\frac{\Gamma_{\textrm{phon}}/2}{\omega_p^2+(\Gamma_{\textrm{phon}}/2)^2}\,.
\end{split}
\label{eq:gamma_pure}
\end{equation}
We investigate the impact of the maximum total number of phonons in the system $N_{\textrm{ph}}$ by calculating linear absorption spectra using different $N_{\textrm{ph}}$ and comparing the respective polaron shifts to the analytic result (\ref{eq:polaron_shift}). As shown in Appendix~\ref{app:convergence}, we find that $N_{\textrm{ph}}=3$ yields very good convergence of $\Delta_{\textrm{pol}}$. We assume that the photon indistinguishability $\mathcal{I}$ is also sufficiently converged in this case. Numerical results for $\mathcal{I}$ in the presence of a cavity can be further corroborated by comparing them to numerically exact results from the (augmented) IBM, which correspond to the limiting case $g=0$.
As shown in Appendix~\ref{app:IBM}, we obtain in this case:
\begin{equation}
 \begin{split}
\mathcal{I}_{g=0}&=\Gamma\int_0^{\infty} d\tau e^{-(\Gamma\tau+\gamma_{\textrm{pure}}\tau-2\textrm{Re}\,\Phi(\tau))}=\Gamma\int_0^{\infty} d\tau e^{-\Gamma\tau}f(\tau)
\end{split}
\label{eq:I_IBM}
\end{equation}
with the phonon dephasing integral (at zero temperature)
\begin{equation}
    \begin{split}
        \Phi(\tau) &= \sum_{\bq} g_{\textrm{ac},\bq}^2\frac{e^{-i(\omega_{\bq}-i\gamma_{\bq}/2)\tau}-1}{(\omega_{\bq}-i\gamma_{\bq}/2)^2} \\
        &= \sum_{p} \frac{M_{p}^2}{\hbar^2}\frac{e^{-i(\omega_p-i\Gamma_{\textrm{phon}}/2)\tau}-1}{(\omega_p-i\Gamma_{\textrm{phon}}/2)^2} \,.
    \end{split}
\label{eq:Phi_tau_final}
\end{equation}
The function $f(\tau)=e^{-(\gamma_{\textrm{pure}}\tau-2\textrm{Re}\,\Phi(\tau))}$ describes the phonon-induced dependence of the HOM interference on the difference time $\tau$.

%\newpage
\section{Results}\label{sec:results}
\begin{figure}
\centering
\includegraphics[width=\columnwidth]{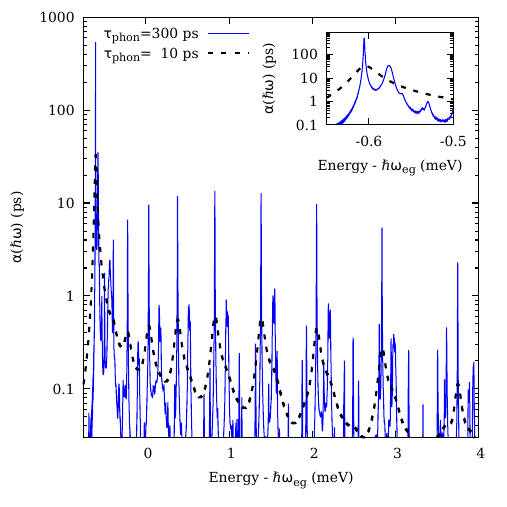}
\caption{Linear absorption spectrum of a WSe$_2$ QD coupled to 2D LA phonons at $T=0$ K for different phonon lifetimes $\tau_{\textrm{phon}}$. Zero energy corresponds to the unperturbed QD transition $\hbar\omega_{\textrm{eg}}$. The inset shows a zoom of the polaron-shifted ZPL.}
\label{fig:spectrum}
\end{figure}

As a first step, we characterize the coupled QD-phonon system by calculating optical absorption spectra in the absence of a cavity for different phonon lifetimes via Eq.~(\ref{eq:spectrum2})
using the unexcited QD as initial state.
%We evaluate the spectrum at time $t=10$ ps without loss of generality. One can show that $\alpha(\omega)$ does not depend on time $t$ in the absence of a cavity, since radiative decay of the QD with the rate $\Gamma$ only introduces a trivial exponential $t$-dependence that cancels in the normalized spectrum. 
The results are shown in Fig.~\ref{fig:spectrum}. The spectrum exhibits a ZPL that is red-shifted with respect to the unperturbed emitter transition $\hbar\omega_{\textrm{eg}}$ due to polaron effects. We extract a polaron shift $\hbar\Delta_{\textrm{pol}} = -0.604$ meV for $\tau_{\textrm{phon}}=10$ ps and $\hbar\Delta_{\textrm{pol}}=-0.606$ meV for $\tau_{\textrm{phon}}=300$ ps from the spectrum, which is close to the numerically exact result $\hbar\Delta_{\textrm{pol}} = -0.611$ meV (for $\tau_{\textrm{phon}}=300$ ps) obtained from the modified IBM (Eq.~(\ref{eq:polaron_shift})).
The phonon sideband (PSB), which is caused by spontaneous phonon emission, appears on the high-energy side of the ZPL in absorption spectra. There is no corresponding sideband at low energies at $T=0$ K. Note that the PSB appears as a series of separate peaks due to the discretization of the phonon mode continuum. Phonon decay leads to a pure dephasing contribution, which increases with decreasing phonon lifetime. It directly leads to a broadening of the ZPL as well as the PSB resonances. This behavior can be understood from the modified IBM result, see Eq.~(\ref{eq:gamma_pure}).

For the following simulations of HOM experiments, we assume that the cavity is tuned into resonance with the polaron-shifted emitter. We therefore use the polaron shift obtained from the absorption spectrum to set $\Delta=\omega_{\textrm{eg}}-\omega_{\textrm{cav}} = -\Delta_{\textrm{pol}}$.
We consider cavities with three different Q-factors $500$, $1500$ and $5000$. According to Eq.~(\ref{eq:Q}), these correspond to $\kappa=4.8$ ps$^{-1}$, $\kappa=1.6$ ps$^{-1}$, and $\kappa=0.48$ ps$^{-1}$, respectively.
\begin{figure}
\centering
\includegraphics[width=1.\columnwidth]{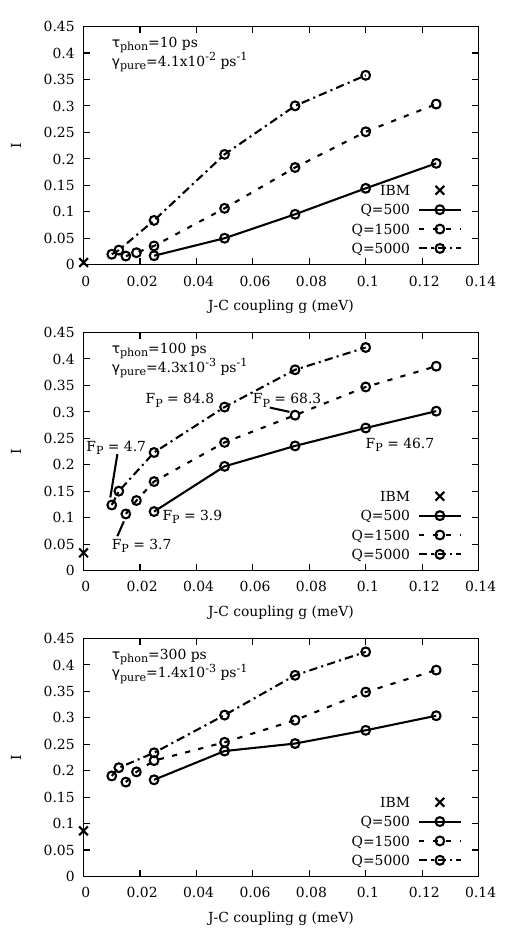}
\caption{Photon indistinguishability $\mathcal{I}$ of a WSe$_2$ QD coupled to 2D LA phonons at $T=0$ K depending on the Jaynes-Cummings (J-C) coupling parameter for different phonon lifetimes $\tau_{\textrm{phon}}$ and different cavity Q-factors. The results are computed by a full solution of the vNL equation. For comparison, we provide results obtained from the augmented independent boson model (IBM), which corresponds to the absence of a cavity. The pure dephasing rate $\gamma_{\textrm{pure}}$ is given for each $\tau_{\textrm{phon}}$ as well as the Purcell factors $F_{\textrm{P}}$ for some of the cavity parameters.}
\label{fig:I_vs_g}
\end{figure}
The dependence of the photon indistinguishability on the phonon lifetime is studied by considering three different values $\tau_{\textrm{phon}}=10$ ps, $\tau_{\textrm{phon}}=100$ ps, and $\tau_{\textrm{phon}}=300$ ps. The corresponding phonon mean free path $\Lambda = c_{\textrm{ac}} \tau_{\textrm{phon}}$ is $40$ nm, $400$ nm, and $1200$ nm, respectively. We compute $\mathcal{I}$ from Eq.~(\ref{eq:I}) using the excited QD without cavity photons as initial state. The results are collected in Fig.~\ref{fig:I_vs_g}. For guidance on the used cavity parameters, we provide the Purcell factors for some of the parameter sets at $\tau_{\textrm{phon}}=100$ ps in the weak- and intermediate-coupling regime. For weak coupling, we obtain $F_{\textrm{P}} \approx 4$ and quadratic scaling $\propto g^2$ is expected. The Purcell factors are calculated as $F_{\textrm{P}} = \tau_{\textrm{free}}/ \tau_{\textrm{cav}}$, where the radiative QD lifetimes $\tau_{\textrm{cav}}$ are obtained via exponential fits to the density-matrix element $\rho^{1,1}_{\textrm{JC}}(t)$, and $\tau_{\textrm{free}}$ enters as a parameter via Eq.~\eqref{eq:tau_free}. We note that although $\gamma_{\textrm{pure}}$ (alongside $\kappa$ and $\Gamma$) in principle influences $F_{\textrm{P}}$, the dephasing is dominated by $\kappa$. We therefore assume that $F_{\textrm{P}}$ only weakly depends on $\tau_{\textrm{phon}}$, so that the provided Purcell factors might be approximately used for the corresponding data points in the upper and lower panels of Fig.~\ref{fig:I_vs_g} as well.
In comparison to single-photon emitters embedded in a bulk material with comparable phonon coupling parameters, such as GaAs \cite{kaer_microscopic_2013}, $\mathcal{I}$ is strongly reduced in the case of WSe$_2$. While near-unity indistinguishability is observed for bulk GaAs, we obtain values between $0.01$ and $0.4$. The main reason for this is the different phonon DOS that leads to more efficient phonon-induced dephasing in the 2D case, as hinted in \cite{mitryakhin_engineering_2024} and exactly verified in this study. 

Despite this fundamental increase in dephasing efficiency, we note that for infinite phonon lifetime, exciton-phonon coupling does not lead to a pure dephasing, i.e. an exponential decay of the HOM interference (\ref{eq:I_IBM}) with difference time $\tau$.
It is straightforward to show that the integrand of $\textrm{Re}\,\Phi(\tau)$, with the dephasing integral given by Eq.~(\ref{eq:Phi_tau_final}),
%the real part of the dephasing integral 
%(\ref{eq:Phi_tau_final}) 
asymptotically scales as $-q^{d-2}\textrm{sin}^2(\omega_q \tau)$ for $d$-dimensional LA phonons and $\gamma_{\bq}=0$. For finite temperatures $T > 0$ K, an additional divergence due to a Bose occupation function $N_q\propto (\omega_q)^{-1}$ would enter, see Eq.~(\ref{eq:Phi_tau_full}). As shown by Krummheuer et al. \cite{krummheuer_theory_2002}, an exponential decay of $\textrm{Re}\,\Phi(\tau)$ would at least require a divergence $\propto -q^{-2} \textrm{sin}^2(\omega_q \tau)$, as it appears in 1D systems at $T>0$ K \cite{ferreira_neto_one-dimensional_2024}. 
Formally, in 2D pure dephasing arises from a quadratic coupling term in the Hamiltonian for the emitter-LA-phonon interaction \cite{vannucci_single-photon_2024}, which we do not consider here. 
%Alternatively, for 2-d systems quadratic coupling to LA phonons needs to be evoked to obtain pure dephasing. 
In our case, pure dephasing automatically results from finite phonon lifetimes, see Eqs.~(\ref{eq:gamma_pure}) and (\ref{eq:I_IBM}).

\begin{figure}
\centering
\includegraphics[width=1.\columnwidth]{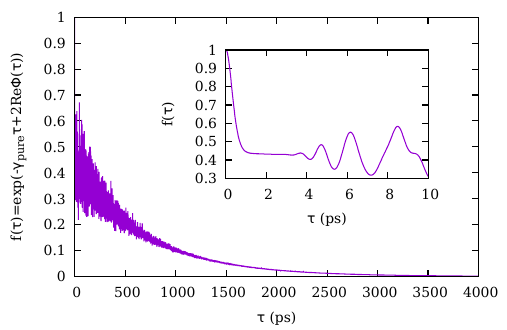}
\caption{Phonon-induced time dependence of the HOM interference $f(\tau)=e^{-(\gamma_{\textrm{pure}}\tau-2\textrm{Re}\,\Phi(\tau))}$ for $\tau_{\textrm{phon}}=300$ ps. The inset shows a zoom-in for very short time scales, where non-Markovian effects due to the PSB become visible.}
\label{fig:exp_Phi_tau}
\end{figure}

Fig.~\ref{fig:I_vs_g} shows how $\mathcal{I}$ is reduced with decreasing phonon lifetime $\tau_{\textrm{phon}}$ (corresponding to increasing $\gamma_{\textrm{pure}}$) when going from bottom to top.
This trend can be understood in a simple picture, where exciton-phonon coupling is only described by pure dephasing, i.e. in the absence of non-Markovian effects, and where the emitter is weakly coupled to a bad cavity $(2g\ll \kappa+\Gamma+\gamma_{\textrm{pure}}, \kappa\gg\Gamma+\gamma_{\textrm{pure}})$. In this case, $\mathcal{I}$ is determined by the effective radiative coupling rate $\Gamma_{\textrm{eff}}$ and $\gamma_{\textrm{pure}}$ alone \cite{grange_cavity-funneled_2015}:
\begin{equation}
    \begin{split}
        \mathcal{I}_{\textrm{simple}}=\frac{\Gamma_{\textrm{eff}}}{\Gamma_{\textrm{eff}}+\gamma_{\textrm{pure}}},\quad \Gamma_{\textrm{eff}}=\Gamma+\frac{4g^2}{\kappa+\Gamma+\gamma_{\textrm{pure}}} \,.
    \end{split}
\label{eq:I_simple}
\end{equation}
For given radiative coupling, an increase of $\gamma_{\textrm{pure}}$ is detrimental for $\mathcal{I}$.
On the other hand, with increasing Jaynes-Cummings coupling $g$ or increasing cavity-Q, the cavity leads to an effectively faster radiative decay, so that $\mathcal{I}$ becomes larger. 
Although an overall trend of increasing $\mathcal{I}$ with increasing $g$ or $Q$ is also visible in Fig.~\ref{fig:I_vs_g}, the curves clearly deviate from the simple behavior suggested by Eq.~(\ref{eq:I_simple}). This is in particular the case for large $\tau_{\textrm{phon}}$, or small $\gamma_{\textrm{pure}}$, where non-Markovian effects of exciton-phonon coupling become more important. Non-Markovian effects are contained in the full quantum dynamics of the vNL equation (\ref{eq:vNL}) or, in the $g=0$-limit, in the dephasing integral $\Phi(\tau)$ of the augmented IBM. To visualize the impact of these effects, we show the phonon-induced dependence of the HOM interference $f(\tau)$ defined in Eq.~(\ref{eq:I_IBM}) for $\tau_{\textrm{phon}}=300$ ps in Fig.~\ref{fig:exp_Phi_tau}. 
We find a fast decay of interference at very short difference times $\tau \approx 1$ ps, which is due to the PSB, and a slow decay at long times corresponding to ZPL broadening via pure dephasing \cite{krummheuer_theory_2002,zimmermann_dephasing_2002}. In the Markovian regime, only the latter effect is included.
As discussed above, due to the $q$-scaling of phonon-induced dephasing, no exponential long-time decay is contained in $e^{2\textrm{Re}\,\Phi(\tau)}$.
Due to these non-Markovian effects, the dependence of $\mathcal{I}$ on the Jaynes-Cummings coupling $g$ is much more complicated than estimated from a pure dephasing model. The importance of non-Markovian effects for $\mathcal{I}$ have also been pointed out for QD in GaAs bulk \cite{kaer_microscopic_2013} and nanowire \cite{ferreira_neto_one-dimensional_2024} systems.

\begin{figure}
\centering
\includegraphics[width=1.\columnwidth]{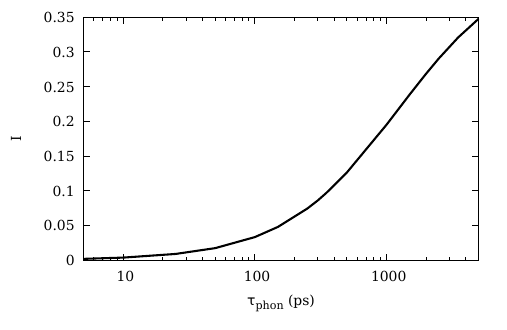}
\caption{Photon indistinguishability $\mathcal{I}$ of a WSe$_2$ QD coupled to 2D LA phonons at $T=0$ K in the absence of a cavity depending on the phonon lifetime $\tau_{\textrm{phon}}$.}
\label{fig:I_g_0}
\end{figure}

Finally, we discuss the dependence of $\mathcal{I}$ on the phonon lifetime in more detail. To this end, we compute $\mathcal{I}$ using the augmented IBM formula (\ref{eq:I_IBM}) in the absence of a cavity for $\tau_{\textrm{phon}}$ varying over several orders of magnitude. The results are shown in Fig.~\ref{fig:I_g_0}.
When the phonon lifetime is extended beyond the values that we have studied in Fig.~\ref{fig:I_vs_g}, indistinguishabilities of more than $0.3$ can be reached even without the Purcell enhancement of a cavity. For $\tau_{\textrm{phon}}\approx 1000$ ps, corresponding to a phonon mean free path $\Lambda\approx 4$ $\mu$m, $\gamma_{\textrm{pure}}$ becomes comparable to the free radiative decay rate $\Gamma=1/2800$ ps$^{-1}$. Hence, in this range, $\mathcal{I}$ starts to saturate, as pure dephasing affects the coherence of emitted single photons less and less. Then, the remaining limitation of indistinguishability is only the non-Markovian dephasing due to the PSB.

\section{Conclusion}

A fully quantum-mechanical description of a WSe$_2$ QD coupled to a cavity as well as to 2D LA phonons is introduced  to quantify the achievable photon indistinguishability $\mathcal{I}$ in a HOM setup. 
Photon indistinguishability in 2D QE systems is strongly reduced compared to semiconductor QD sources embedded in bulk material due to more efficient phonon-induced dephasing, which can be traced back to the dimensionality of the phonon DOS.
Even more, phonon decay has a detrimental effect on $\mathcal{I}$ by causing pure dephasing. At low temperatures, the phonon lifetime in turn depends on the mean free path of phonons in the sample given by the distance of the QD emitter to grain boundaries or the sample surface. Therefore, sample quality and size have a direct impact on photon indistinguishability.

We find a beneficial effect of embedding the 2D QEs in a cavity, as $\mathcal{I}$ generally increases with increasing light-matter coupling $g$ and cavity-Q factor, as expected from a Markovian weak-coupling picture. However, strong non-Markovian effects due to the PSB lead to a nontrivial dependence on $g$.
We find a maximum achievable $\mathcal{I}$ of $0.3$ to $0.4$ in these QD-cavity systems under standard excitation conditions and at $T=0$ K. At finite temperatures, additional exciton-phonon scattering channels are expected to further reduce coherence.
Addressing the issue of phonon-induced detrimental effects in single-photon sources made from vdW semiconductors will be a major challenge in the 2D materials community. One possible route could be to use phonon engineering as a tool to decouple the QD from the phonons, e.g. in nanostructure superlattices \cite{ortiz_phonon_2019}.

%and a route towards becoming competitive with the well-established III/V material platform.

\newpage
\textbf{Acknowledgement}
The project is funded by the Deutsche Forschungsgemeinschaft (DFG) in the priority program SPP 2244 (funding numbers:
Schn1376/14.2 and Gi1121/4-2) and by the QuantERA II European Union’s Horizon 2020 research and innovation program under the EQUAISE project (Grant Agreement No.101017733).
Funding by the BMBF within the project TUBLAN and the project QR.N is acknowledged.
We gratefully acknowledge the computing time provided on the supercomputer Emmy/Grete at NHR@Göttingen as part of the NHR infrastructure.

% \bibliography{HOM, HOM_2025_IS,2025_HOM_Cavities_2D}

\begin{thebibliography}{54}%
\makeatletter
\providecommand \@ifxundefined [1]{%
 \@ifx{#1\undefined}
}%
\providecommand \@ifnum [1]{%
 \ifnum #1\expandafter \@firstoftwo
 \else \expandafter \@secondoftwo
 \fi
}%
\providecommand \@ifx [1]{%
 \ifx #1\expandafter \@firstoftwo
 \else \expandafter \@secondoftwo
 \fi
}%
\providecommand \natexlab [1]{#1}%
\providecommand \enquote  [1]{``#1''}%
\providecommand \bibnamefont  [1]{#1}%
\providecommand \bibfnamefont [1]{#1}%
\providecommand \citenamefont [1]{#1}%
\providecommand \href@noop [0]{\@secondoftwo}%
\providecommand \href [0]{\begingroup \@sanitize@url \@href}%
\providecommand \@href[1]{\@@startlink{#1}\@@href}%
\providecommand \@@href[1]{\endgroup#1\@@endlink}%
\providecommand \@sanitize@url [0]{\catcode `\\12\catcode `\$12\catcode
  `\&12\catcode `\#12\catcode `\^12\catcode `\_12\catcode `\%12\relax}%
\providecommand \@@startlink[1]{}%
\providecommand \@@endlink[0]{}%
\providecommand \url  [0]{\begingroup\@sanitize@url \@url }%
\providecommand \@url [1]{\endgroup\@href {#1}{\urlprefix }}%
\providecommand \urlprefix  [0]{URL }%
\providecommand \Eprint [0]{\href }%
\providecommand \doibase [0]{https://doi.org/}%
\providecommand \selectlanguage [0]{\@gobble}%
\providecommand \bibinfo  [0]{\@secondoftwo}%
\providecommand \bibfield  [0]{\@secondoftwo}%
\providecommand \translation [1]{[#1]}%
\providecommand \BibitemOpen [0]{}%
\providecommand \bibitemStop [0]{}%
\providecommand \bibitemNoStop [0]{.\EOS\space}%
\providecommand \EOS [0]{\spacefactor3000\relax}%
\providecommand \BibitemShut  [1]{\csname bibitem#1\endcsname}%
\let\auto@bib@innerbib\@empty
%</preamble>
\bibitem [{\citenamefont {Kimble}(2008)}]{kimble_quantum_2008}%
  \BibitemOpen
  \bibfield  {author} {\bibinfo {author} {\bibfnamefont {H.~J.}\ \bibnamefont
  {Kimble}},\ }\bibfield  {title} {\bibinfo {title} {The quantum internet},\
  }\href {https://doi.org/10.1038/nature07127} {\bibfield  {journal} {\bibinfo
  {journal} {Nature}\ }\textbf {\bibinfo {volume} {453}},\ \bibinfo {pages}
  {1023} (\bibinfo {year} {2008})},\ \bibinfo {note} {publisher: Nature
  Publishing Group}\BibitemShut {NoStop}%
\bibitem [{\citenamefont {Aharonovich}\ \emph {et~al.}(2016)\citenamefont
  {Aharonovich}, \citenamefont {Englund},\ and\ \citenamefont
  {Toth}}]{aharonovich_solid-state_2016}%
  \BibitemOpen
  \bibfield  {author} {\bibinfo {author} {\bibfnamefont {I.}~\bibnamefont
  {Aharonovich}}, \bibinfo {author} {\bibfnamefont {D.}~\bibnamefont
  {Englund}},\ and\ \bibinfo {author} {\bibfnamefont {M.}~\bibnamefont
  {Toth}},\ }\bibfield  {title} {\bibinfo {title} {Solid-state single-photon
  emitters},\ }\href {https://doi.org/10.1038/nphoton.2016.186} {\bibfield
  {journal} {\bibinfo  {journal} {Nature Photon}\ }\textbf {\bibinfo {volume}
  {10}},\ \bibinfo {pages} {631} (\bibinfo {year} {2016})},\ \bibinfo {note}
  {publisher: Nature Publishing Group}\BibitemShut {NoStop}%
\bibitem [{\citenamefont {Gao}\ \emph {et~al.}(2023)\citenamefont {Gao},
  \citenamefont {Von~Helversen}, \citenamefont {Antón-Solanas}, \citenamefont
  {Schneider},\ and\ \citenamefont {Heindel}}]{gao_atomically-thin_2023}%
  \BibitemOpen
  \bibfield  {author} {\bibinfo {author} {\bibfnamefont {T.}~\bibnamefont
  {Gao}}, \bibinfo {author} {\bibfnamefont {M.}~\bibnamefont {Von~Helversen}},
  \bibinfo {author} {\bibfnamefont {C.}~\bibnamefont {Antón-Solanas}},
  \bibinfo {author} {\bibfnamefont {C.}~\bibnamefont {Schneider}},\ and\
  \bibinfo {author} {\bibfnamefont {T.}~\bibnamefont {Heindel}},\ }\bibfield
  {title} {\bibinfo {title} {Atomically-thin single-photon sources for quantum
  communication},\ }\href {https://doi.org/10.1038/s41699-023-00366-4}
  {\bibfield  {journal} {\bibinfo  {journal} {npj 2D Mater Appl}\ }\textbf
  {\bibinfo {volume} {7}},\ \bibinfo {pages} {4} (\bibinfo {year}
  {2023})}\BibitemShut {NoStop}%
\bibitem [{\citenamefont {Flissikowski}\ \emph {et~al.}(2004)\citenamefont
  {Flissikowski}, \citenamefont {Betke}, \citenamefont {Akimov},\ and\
  \citenamefont {Henneberger}}]{flissikowski_two-photon_2004}%
  \BibitemOpen
  \bibfield  {author} {\bibinfo {author} {\bibfnamefont {T.}~\bibnamefont
  {Flissikowski}}, \bibinfo {author} {\bibfnamefont {A.}~\bibnamefont {Betke}},
  \bibinfo {author} {\bibfnamefont {I.~A.}\ \bibnamefont {Akimov}},\ and\
  \bibinfo {author} {\bibfnamefont {F.}~\bibnamefont {Henneberger}},\
  }\bibfield  {title} {\bibinfo {title} {Two-{Photon} {Coherent} {Control} of a
  {Single} {Quantum} {Dot}},\ }\href
  {https://doi.org/10.1103/PhysRevLett.92.227401} {\bibfield  {journal}
  {\bibinfo  {journal} {Phys. Rev. Lett.}\ }\textbf {\bibinfo {volume} {92}},\
  \bibinfo {pages} {227401} (\bibinfo {year} {2004})},\ \bibinfo {note}
  {publisher: American Physical Society}\BibitemShut {NoStop}%
\bibitem [{\citenamefont {Jayakumar}\ \emph {et~al.}(2013)\citenamefont
  {Jayakumar}, \citenamefont {Predojević}, \citenamefont {Huber},
  \citenamefont {Kauten}, \citenamefont {Solomon},\ and\ \citenamefont
  {Weihs}}]{jayakumar_deterministic_2013}%
  \BibitemOpen
  \bibfield  {author} {\bibinfo {author} {\bibfnamefont {H.}~\bibnamefont
  {Jayakumar}}, \bibinfo {author} {\bibfnamefont {A.}~\bibnamefont
  {Predojević}}, \bibinfo {author} {\bibfnamefont {T.}~\bibnamefont {Huber}},
  \bibinfo {author} {\bibfnamefont {T.}~\bibnamefont {Kauten}}, \bibinfo
  {author} {\bibfnamefont {G.~S.}\ \bibnamefont {Solomon}},\ and\ \bibinfo
  {author} {\bibfnamefont {G.}~\bibnamefont {Weihs}},\ }\bibfield  {title}
  {\bibinfo {title} {Deterministic {Photon} {Pairs} and {Coherent} {Optical}
  {Control} of a {Single} {Quantum} {Dot}},\ }\href
  {https://doi.org/10.1103/PhysRevLett.110.135505} {\bibfield  {journal}
  {\bibinfo  {journal} {Phys. Rev. Lett.}\ }\textbf {\bibinfo {volume} {110}},\
  \bibinfo {pages} {135505} (\bibinfo {year} {2013})},\ \bibinfo {note}
  {publisher: American Physical Society}\BibitemShut {NoStop}%
\bibitem [{\citenamefont {Ma}\ \emph {et~al.}(2016)\citenamefont {Ma},
  \citenamefont {Yuan}, \citenamefont {Cao}, \citenamefont {Zhang},\ and\
  \citenamefont {Qi}}]{ma_quantum_2016}%
  \BibitemOpen
  \bibfield  {author} {\bibinfo {author} {\bibfnamefont {X.}~\bibnamefont
  {Ma}}, \bibinfo {author} {\bibfnamefont {X.}~\bibnamefont {Yuan}}, \bibinfo
  {author} {\bibfnamefont {Z.}~\bibnamefont {Cao}}, \bibinfo {author}
  {\bibfnamefont {Z.}~\bibnamefont {Zhang}},\ and\ \bibinfo {author}
  {\bibfnamefont {B.}~\bibnamefont {Qi}},\ }\bibfield  {title}
  {{\selectlanguage {English}\bibinfo {title} {Quantum random number
  generation}},\ }\bibfield  {journal} {\bibinfo  {journal} {npj Quantum
  Information}\ }\textbf {\bibinfo {volume} {2}},\ \href
  {https://doi.org/10.1038/npjqi.2016.21} {10.1038/npjqi.2016.21} (\bibinfo
  {year} {2016}),\ \bibinfo {note} {institution: Oak Ridge National Laboratory
  (ORNL), Oak Ridge, TN (United States) Publisher: Nature Partner
  Journals}\BibitemShut {NoStop}%
\bibitem [{\citenamefont {O'Brien}\ \emph {et~al.}(2009)\citenamefont
  {O'Brien}, \citenamefont {Furusawa},\ and\ \citenamefont
  {Vučković}}]{obrien_photonic_2009}%
  \BibitemOpen
  \bibfield  {author} {\bibinfo {author} {\bibfnamefont {J.~L.}\ \bibnamefont
  {O'Brien}}, \bibinfo {author} {\bibfnamefont {A.}~\bibnamefont {Furusawa}},\
  and\ \bibinfo {author} {\bibfnamefont {J.}~\bibnamefont {Vučković}},\
  }\bibfield  {title} {\bibinfo {title} {Photonic quantum technologies},\
  }\href {https://doi.org/10.1038/nphoton.2009.229} {\bibfield  {journal}
  {\bibinfo  {journal} {Nature Photon}\ }\textbf {\bibinfo {volume} {3}},\
  \bibinfo {pages} {687} (\bibinfo {year} {2009})},\ \bibinfo {note}
  {publisher: Nature Publishing Group}\BibitemShut {NoStop}%
\bibitem [{\citenamefont {Wang}\ \emph
  {et~al.}(2019{\natexlab{a}})\citenamefont {Wang}, \citenamefont {Qin},
  \citenamefont {Ding}, \citenamefont {Chen}, \citenamefont {Chen},
  \citenamefont {You}, \citenamefont {He}, \citenamefont {Jiang}, \citenamefont
  {You}, \citenamefont {Wang}, \citenamefont {Schneider}, \citenamefont
  {Renema}, \citenamefont {Höfling}, \citenamefont {Lu},\ and\ \citenamefont
  {Pan}}]{wang_boson_2019}%
  \BibitemOpen
  \bibfield  {author} {\bibinfo {author} {\bibfnamefont {H.}~\bibnamefont
  {Wang}}, \bibinfo {author} {\bibfnamefont {J.}~\bibnamefont {Qin}}, \bibinfo
  {author} {\bibfnamefont {X.}~\bibnamefont {Ding}}, \bibinfo {author}
  {\bibfnamefont {M.-C.}\ \bibnamefont {Chen}}, \bibinfo {author}
  {\bibfnamefont {S.}~\bibnamefont {Chen}}, \bibinfo {author} {\bibfnamefont
  {X.}~\bibnamefont {You}}, \bibinfo {author} {\bibfnamefont {Y.-M.}\
  \bibnamefont {He}}, \bibinfo {author} {\bibfnamefont {X.}~\bibnamefont
  {Jiang}}, \bibinfo {author} {\bibfnamefont {L.}~\bibnamefont {You}}, \bibinfo
  {author} {\bibfnamefont {Z.}~\bibnamefont {Wang}}, \bibinfo {author}
  {\bibfnamefont {C.}~\bibnamefont {Schneider}}, \bibinfo {author}
  {\bibfnamefont {J.~J.}\ \bibnamefont {Renema}}, \bibinfo {author}
  {\bibfnamefont {S.}~\bibnamefont {Höfling}}, \bibinfo {author}
  {\bibfnamefont {C.-Y.}\ \bibnamefont {Lu}},\ and\ \bibinfo {author}
  {\bibfnamefont {J.-W.}\ \bibnamefont {Pan}},\ }\bibfield  {title} {\bibinfo
  {title} {Boson sampling with 20 input photons and a 60-mode interferometer in
  a $1{0}^{14}$-dimensional hilbert space},\ }\href
  {https://doi.org/10.1103/PhysRevLett.123.250503} {\bibfield  {journal}
  {\bibinfo  {journal} {Physical Review Letters}\ }\textbf {\bibinfo {volume}
  {123}},\ \bibinfo {pages} {250503} (\bibinfo {year} {2019}{\natexlab{a}})},\
  \bibinfo {note} {publisher: American Physical Society}\BibitemShut {NoStop}%
\bibitem [{\citenamefont {Lal}\ \emph {et~al.}(2022)\citenamefont {Lal},
  \citenamefont {Mishra},\ and\ \citenamefont
  {Singh}}]{lal_indistinguishable_2022}%
  \BibitemOpen
  \bibfield  {author} {\bibinfo {author} {\bibfnamefont {N.}~\bibnamefont
  {Lal}}, \bibinfo {author} {\bibfnamefont {S.}~\bibnamefont {Mishra}},\ and\
  \bibinfo {author} {\bibfnamefont {R.~P.}\ \bibnamefont {Singh}},\ }\bibfield
  {title} {\bibinfo {title} {Indistinguishable photons},\ }\href
  {https://doi.org/10.1116/5.0083968} {\bibfield  {journal} {\bibinfo
  {journal} {{AVS} Quantum Science}\ }\textbf {\bibinfo {volume} {4}},\
  \bibinfo {pages} {021701} (\bibinfo {year} {2022})}\BibitemShut {NoStop}%
\bibitem [{\citenamefont {Peyskens}\ \emph {et~al.}(2019)\citenamefont
  {Peyskens}, \citenamefont {Chakraborty}, \citenamefont {Muneeb},
  \citenamefont {Van~Thourhout},\ and\ \citenamefont
  {Englund}}]{peyskens_integration_2019}%
  \BibitemOpen
  \bibfield  {author} {\bibinfo {author} {\bibfnamefont {F.}~\bibnamefont
  {Peyskens}}, \bibinfo {author} {\bibfnamefont {C.}~\bibnamefont
  {Chakraborty}}, \bibinfo {author} {\bibfnamefont {M.}~\bibnamefont {Muneeb}},
  \bibinfo {author} {\bibfnamefont {D.}~\bibnamefont {Van~Thourhout}},\ and\
  \bibinfo {author} {\bibfnamefont {D.}~\bibnamefont {Englund}},\ }\bibfield
  {title} {\bibinfo {title} {Integration of single photon emitters in {2D}
  layered materials with a silicon nitride photonic chip},\ }\href
  {https://doi.org/10.1038/s41467-019-12421-0} {\bibfield  {journal} {\bibinfo
  {journal} {Nature Communications}\ }\textbf {\bibinfo {volume} {10}},\
  \bibinfo {pages} {4435} (\bibinfo {year} {2019})},\ \bibinfo {note}
  {publisher: Nature Publishing Group}\BibitemShut {NoStop}%
\bibitem [{\citenamefont {Hong}\ \emph {et~al.}(1987)\citenamefont {Hong},
  \citenamefont {Ou},\ and\ \citenamefont {Mandel}}]{hong_measurement_1987}%
  \BibitemOpen
  \bibfield  {author} {\bibinfo {author} {\bibfnamefont {C.~K.}\ \bibnamefont
  {Hong}}, \bibinfo {author} {\bibfnamefont {Z.~Y.}\ \bibnamefont {Ou}},\ and\
  \bibinfo {author} {\bibfnamefont {L.}~\bibnamefont {Mandel}},\ }\bibfield
  {title} {\bibinfo {title} {Measurement of subpicosecond time intervals
  between two photons by interference},\ }\href
  {https://doi.org/10.1103/PhysRevLett.59.2044} {\bibfield  {journal} {\bibinfo
   {journal} {Phys. Rev. Lett.}\ }\textbf {\bibinfo {volume} {59}},\ \bibinfo
  {pages} {2044} (\bibinfo {year} {1987})},\ \bibinfo {note} {publisher:
  American Physical Society}\BibitemShut {NoStop}%
\bibitem [{\citenamefont {Santori}\ \emph {et~al.}(2002)\citenamefont
  {Santori}, \citenamefont {Fattal}, \citenamefont {Vučković}, \citenamefont
  {Solomon},\ and\ \citenamefont {Yamamoto}}]{santori_indistinguishable_2002}%
  \BibitemOpen
  \bibfield  {author} {\bibinfo {author} {\bibfnamefont {C.}~\bibnamefont
  {Santori}}, \bibinfo {author} {\bibfnamefont {D.}~\bibnamefont {Fattal}},
  \bibinfo {author} {\bibfnamefont {J.}~\bibnamefont {Vučković}}, \bibinfo
  {author} {\bibfnamefont {G.~S.}\ \bibnamefont {Solomon}},\ and\ \bibinfo
  {author} {\bibfnamefont {Y.}~\bibnamefont {Yamamoto}},\ }\bibfield  {title}
  {\bibinfo {title} {Indistinguishable photons from a single-photon device},\
  }\href {https://doi.org/10.1038/nature01086} {\bibfield  {journal} {\bibinfo
  {journal} {Nature}\ }\textbf {\bibinfo {volume} {419}},\ \bibinfo {pages}
  {594} (\bibinfo {year} {2002})},\ \bibinfo {note} {publisher: Nature
  Publishing Group}\BibitemShut {NoStop}%
\bibitem [{\citenamefont {Wang}\ \emph
  {et~al.}(2019{\natexlab{b}})\citenamefont {Wang}, \citenamefont {He},
  \citenamefont {Chung}, \citenamefont {Hu}, \citenamefont {Yu}, \citenamefont
  {Chen}, \citenamefont {Ding}, \citenamefont {Chen}, \citenamefont {Qin},
  \citenamefont {Yang}, \citenamefont {Liu}, \citenamefont {Duan},
  \citenamefont {Li}, \citenamefont {Gerhardt}, \citenamefont {Winkler},
  \citenamefont {Jurkat}, \citenamefont {Wang}, \citenamefont {Gregersen},
  \citenamefont {Huo}, \citenamefont {Dai}, \citenamefont {Yu}, \citenamefont
  {Höfling}, \citenamefont {Lu},\ and\ \citenamefont
  {Pan}}]{wang_towards_2019}%
  \BibitemOpen
  \bibfield  {author} {\bibinfo {author} {\bibfnamefont {H.}~\bibnamefont
  {Wang}}, \bibinfo {author} {\bibfnamefont {Y.-M.}\ \bibnamefont {He}},
  \bibinfo {author} {\bibfnamefont {T.-H.}\ \bibnamefont {Chung}}, \bibinfo
  {author} {\bibfnamefont {H.}~\bibnamefont {Hu}}, \bibinfo {author}
  {\bibfnamefont {Y.}~\bibnamefont {Yu}}, \bibinfo {author} {\bibfnamefont
  {S.}~\bibnamefont {Chen}}, \bibinfo {author} {\bibfnamefont {X.}~\bibnamefont
  {Ding}}, \bibinfo {author} {\bibfnamefont {M.-C.}\ \bibnamefont {Chen}},
  \bibinfo {author} {\bibfnamefont {J.}~\bibnamefont {Qin}}, \bibinfo {author}
  {\bibfnamefont {X.}~\bibnamefont {Yang}}, \bibinfo {author} {\bibfnamefont
  {R.-Z.}\ \bibnamefont {Liu}}, \bibinfo {author} {\bibfnamefont {Z.-C.}\
  \bibnamefont {Duan}}, \bibinfo {author} {\bibfnamefont {J.-P.}\ \bibnamefont
  {Li}}, \bibinfo {author} {\bibfnamefont {S.}~\bibnamefont {Gerhardt}},
  \bibinfo {author} {\bibfnamefont {K.}~\bibnamefont {Winkler}}, \bibinfo
  {author} {\bibfnamefont {J.}~\bibnamefont {Jurkat}}, \bibinfo {author}
  {\bibfnamefont {L.-J.}\ \bibnamefont {Wang}}, \bibinfo {author}
  {\bibfnamefont {N.}~\bibnamefont {Gregersen}}, \bibinfo {author}
  {\bibfnamefont {Y.-H.}\ \bibnamefont {Huo}}, \bibinfo {author} {\bibfnamefont
  {Q.}~\bibnamefont {Dai}}, \bibinfo {author} {\bibfnamefont {S.}~\bibnamefont
  {Yu}}, \bibinfo {author} {\bibfnamefont {S.}~\bibnamefont {Höfling}},
  \bibinfo {author} {\bibfnamefont {C.-Y.}\ \bibnamefont {Lu}},\ and\ \bibinfo
  {author} {\bibfnamefont {J.-W.}\ \bibnamefont {Pan}},\ }\bibfield  {title}
  {\bibinfo {title} {Towards optimal single-photon sources from polarized
  microcavities},\ }\href {https://doi.org/10.1038/s41566-019-0494-3}
  {\bibfield  {journal} {\bibinfo  {journal} {Nat. Photonics}\ }\textbf
  {\bibinfo {volume} {13}},\ \bibinfo {pages} {770} (\bibinfo {year}
  {2019}{\natexlab{b}})},\ \bibinfo {note} {publisher: Nature Publishing
  Group}\BibitemShut {NoStop}%
\bibitem [{\citenamefont {Komza}\ \emph {et~al.}(2024)\citenamefont {Komza},
  \citenamefont {Samutpraphoot}, \citenamefont {Odeh}, \citenamefont {Tang},
  \citenamefont {Mathew}, \citenamefont {Chang}, \citenamefont {Song},
  \citenamefont {Kim}, \citenamefont {Xiong}, \citenamefont {Hautier},\ and\
  \citenamefont {Sipahigil}}]{komza_indistinguishable_2024}%
  \BibitemOpen
  \bibfield  {author} {\bibinfo {author} {\bibfnamefont {L.}~\bibnamefont
  {Komza}}, \bibinfo {author} {\bibfnamefont {P.}~\bibnamefont
  {Samutpraphoot}}, \bibinfo {author} {\bibfnamefont {M.}~\bibnamefont {Odeh}},
  \bibinfo {author} {\bibfnamefont {Y.-L.}\ \bibnamefont {Tang}}, \bibinfo
  {author} {\bibfnamefont {M.}~\bibnamefont {Mathew}}, \bibinfo {author}
  {\bibfnamefont {J.}~\bibnamefont {Chang}}, \bibinfo {author} {\bibfnamefont
  {H.}~\bibnamefont {Song}}, \bibinfo {author} {\bibfnamefont {M.-K.}\
  \bibnamefont {Kim}}, \bibinfo {author} {\bibfnamefont {Y.}~\bibnamefont
  {Xiong}}, \bibinfo {author} {\bibfnamefont {G.}~\bibnamefont {Hautier}},\
  and\ \bibinfo {author} {\bibfnamefont {A.}~\bibnamefont {Sipahigil}},\
  }\bibfield  {title} {\bibinfo {title} {Indistinguishable photons from an
  artificial atom in silicon photonics},\ }\href
  {https://doi.org/10.1038/s41467-024-51265-1} {\bibfield  {journal} {\bibinfo
  {journal} {Nature Communications}\ }\textbf {\bibinfo {volume} {15}},\
  \bibinfo {pages} {6920} (\bibinfo {year} {2024})},\ \bibinfo {note}
  {publisher: Nature Publishing Group}\BibitemShut {NoStop}%
\bibitem [{\citenamefont {Wei}\ \emph {et~al.}(2014)\citenamefont {Wei},
  \citenamefont {He}, \citenamefont {Chen}, \citenamefont {Hu}, \citenamefont
  {He}, \citenamefont {Wu}, \citenamefont {Schneider}, \citenamefont {Kamp},
  \citenamefont {Höfling}, \citenamefont {Lu},\ and\ \citenamefont
  {Pan}}]{wei_deterministic_2014}%
  \BibitemOpen
  \bibfield  {author} {\bibinfo {author} {\bibfnamefont {Y.-J.}\ \bibnamefont
  {Wei}}, \bibinfo {author} {\bibfnamefont {Y.-M.}\ \bibnamefont {He}},
  \bibinfo {author} {\bibfnamefont {M.-C.}\ \bibnamefont {Chen}}, \bibinfo
  {author} {\bibfnamefont {Y.-N.}\ \bibnamefont {Hu}}, \bibinfo {author}
  {\bibfnamefont {Y.}~\bibnamefont {He}}, \bibinfo {author} {\bibfnamefont
  {D.}~\bibnamefont {Wu}}, \bibinfo {author} {\bibfnamefont {C.}~\bibnamefont
  {Schneider}}, \bibinfo {author} {\bibfnamefont {M.}~\bibnamefont {Kamp}},
  \bibinfo {author} {\bibfnamefont {S.}~\bibnamefont {Höfling}}, \bibinfo
  {author} {\bibfnamefont {C.-Y.}\ \bibnamefont {Lu}},\ and\ \bibinfo {author}
  {\bibfnamefont {J.-W.}\ \bibnamefont {Pan}},\ }\bibfield  {title} {\bibinfo
  {title} {Deterministic and {Robust} {Generation} of {Single} {Photons} from a
  {Single} {Quantum} {Dot} with 99.5\% {Indistinguishability} {Using}
  {Adiabatic} {Rapid} {Passage}},\ }\href {https://doi.org/10.1021/nl503081n}
  {\bibfield  {journal} {\bibinfo  {journal} {Nano Lett.}\ }\textbf {\bibinfo
  {volume} {14}},\ \bibinfo {pages} {6515} (\bibinfo {year}
  {2014})}\BibitemShut {NoStop}%
\bibitem [{\citenamefont {Yu}\ \emph {et~al.}(2024)\citenamefont {Yu},
  \citenamefont {Seo}, \citenamefont {Luo}, \citenamefont {Lu}, \citenamefont
  {Son}, \citenamefont {Tan},\ and\ \citenamefont {Nam}}]{yu_tunable_2024}%
  \BibitemOpen
  \bibfield  {author} {\bibinfo {author} {\bibfnamefont {Y.}~\bibnamefont
  {Yu}}, \bibinfo {author} {\bibfnamefont {I.~C.}\ \bibnamefont {Seo}},
  \bibinfo {author} {\bibfnamefont {M.}~\bibnamefont {Luo}}, \bibinfo {author}
  {\bibfnamefont {K.}~\bibnamefont {Lu}}, \bibinfo {author} {\bibfnamefont
  {B.}~\bibnamefont {Son}}, \bibinfo {author} {\bibfnamefont {J.~K.}\
  \bibnamefont {Tan}},\ and\ \bibinfo {author} {\bibfnamefont {D.}~\bibnamefont
  {Nam}},\ }\bibfield  {title} {\bibinfo {title} {Tunable single-photon
  emitters in {2D} materials},\ }\href
  {https://doi.org/10.1515/nanoph-2024-0050} {\bibfield  {journal} {\bibinfo
  {journal} {Nanophotonics}\ }\textbf {\bibinfo {volume} {13}},\ \bibinfo
  {pages} {3615} (\bibinfo {year} {2024})},\ \bibinfo {note} {publisher: De
  Gruyter}\BibitemShut {NoStop}%
\bibitem [{\citenamefont {Kern}\ \emph {et~al.}(2016)\citenamefont {Kern},
  \citenamefont {Niehues}, \citenamefont {Tonndorf}, \citenamefont {Schmidt},
  \citenamefont {Wigger}, \citenamefont {Schneider}, \citenamefont {Stiehm},
  \citenamefont {Michaelis De~Vasconcellos}, \citenamefont {Reiter},
  \citenamefont {Kuhn},\ and\ \citenamefont
  {Bratschitsch}}]{kern_nanoscale_2016}%
  \BibitemOpen
  \bibfield  {author} {\bibinfo {author} {\bibfnamefont {J.}~\bibnamefont
  {Kern}}, \bibinfo {author} {\bibfnamefont {I.}~\bibnamefont {Niehues}},
  \bibinfo {author} {\bibfnamefont {P.}~\bibnamefont {Tonndorf}}, \bibinfo
  {author} {\bibfnamefont {R.}~\bibnamefont {Schmidt}}, \bibinfo {author}
  {\bibfnamefont {D.}~\bibnamefont {Wigger}}, \bibinfo {author} {\bibfnamefont
  {R.}~\bibnamefont {Schneider}}, \bibinfo {author} {\bibfnamefont
  {T.}~\bibnamefont {Stiehm}}, \bibinfo {author} {\bibfnamefont
  {S.}~\bibnamefont {Michaelis De~Vasconcellos}}, \bibinfo {author}
  {\bibfnamefont {D.~E.}\ \bibnamefont {Reiter}}, \bibinfo {author}
  {\bibfnamefont {T.}~\bibnamefont {Kuhn}},\ and\ \bibinfo {author}
  {\bibfnamefont {R.}~\bibnamefont {Bratschitsch}},\ }\bibfield  {title}
  {\bibinfo {title} {Nanoscale {Positioning} of {Single}‐{Photon} {Emitters}
  in {Atomically} {Thin} {WSe}$_{\textrm{2}}$},\ }\href
  {https://doi.org/10.1002/adma.201600560} {\bibfield  {journal} {\bibinfo
  {journal} {Advanced Materials}\ }\textbf {\bibinfo {volume} {28}},\ \bibinfo
  {pages} {7101} (\bibinfo {year} {2016})}\BibitemShut {NoStop}%
\bibitem [{\citenamefont {Palacios-Berraquero}\ \emph
  {et~al.}(2017)\citenamefont {Palacios-Berraquero}, \citenamefont {Kara},
  \citenamefont {Montblanch}, \citenamefont {Barbone}, \citenamefont
  {Latawiec}, \citenamefont {Yoon}, \citenamefont {Ott}, \citenamefont
  {Loncar}, \citenamefont {Ferrari},\ and\ \citenamefont
  {Atatüre}}]{palacios-berraquero_large-scale_2017}%
  \BibitemOpen
  \bibfield  {author} {\bibinfo {author} {\bibfnamefont {C.}~\bibnamefont
  {Palacios-Berraquero}}, \bibinfo {author} {\bibfnamefont {D.~M.}\
  \bibnamefont {Kara}}, \bibinfo {author} {\bibfnamefont {A.~R.-P.}\
  \bibnamefont {Montblanch}}, \bibinfo {author} {\bibfnamefont
  {M.}~\bibnamefont {Barbone}}, \bibinfo {author} {\bibfnamefont
  {P.}~\bibnamefont {Latawiec}}, \bibinfo {author} {\bibfnamefont
  {D.}~\bibnamefont {Yoon}}, \bibinfo {author} {\bibfnamefont {A.~K.}\
  \bibnamefont {Ott}}, \bibinfo {author} {\bibfnamefont {M.}~\bibnamefont
  {Loncar}}, \bibinfo {author} {\bibfnamefont {A.~C.}\ \bibnamefont
  {Ferrari}},\ and\ \bibinfo {author} {\bibfnamefont {M.}~\bibnamefont
  {Atatüre}},\ }\bibfield  {title} {\bibinfo {title} {Large-scale
  quantum-emitter arrays in atomically thin semiconductors},\ }\href
  {https://doi.org/10.1038/ncomms15093} {\bibfield  {journal} {\bibinfo
  {journal} {Nat Commun}\ }\textbf {\bibinfo {volume} {8}},\ \bibinfo {pages}
  {15093} (\bibinfo {year} {2017})}\BibitemShut {NoStop}%
\bibitem [{\citenamefont {Yu}\ \emph {et~al.}(2021)\citenamefont {Yu},
  \citenamefont {Deng}, \citenamefont {Zhang}, \citenamefont {Borghardt},
  \citenamefont {Kardynal}, \citenamefont {Vučković},\ and\ \citenamefont
  {Heinz}}]{yu_site-controlled_2021}%
  \BibitemOpen
  \bibfield  {author} {\bibinfo {author} {\bibfnamefont {L.}~\bibnamefont
  {Yu}}, \bibinfo {author} {\bibfnamefont {M.}~\bibnamefont {Deng}}, \bibinfo
  {author} {\bibfnamefont {J.~L.}\ \bibnamefont {Zhang}}, \bibinfo {author}
  {\bibfnamefont {S.}~\bibnamefont {Borghardt}}, \bibinfo {author}
  {\bibfnamefont {B.}~\bibnamefont {Kardynal}}, \bibinfo {author}
  {\bibfnamefont {J.}~\bibnamefont {Vučković}},\ and\ \bibinfo {author}
  {\bibfnamefont {T.~F.}\ \bibnamefont {Heinz}},\ }\bibfield  {title} {\bibinfo
  {title} {Site-{Controlled} {Quantum} {Emitters} in {Monolayer}
  {MoSe}$_{\textrm{2}}$},\ }\href
  {https://doi.org/10.1021/acs.nanolett.0c04282} {\bibfield  {journal}
  {\bibinfo  {journal} {Nano Lett.}\ }\textbf {\bibinfo {volume} {21}},\
  \bibinfo {pages} {2376} (\bibinfo {year} {2021})}\BibitemShut {NoStop}%
\bibitem [{\citenamefont {Wang}\ \emph {et~al.}(2021)\citenamefont {Wang},
  \citenamefont {Maisch}, \citenamefont {Tang}, \citenamefont {Zhao},
  \citenamefont {Yang}, \citenamefont {Joos}, \citenamefont {Portalupi},
  \citenamefont {Michler},\ and\ \citenamefont {Smet}}]{wang_highly_2021}%
  \BibitemOpen
  \bibfield  {author} {\bibinfo {author} {\bibfnamefont {Q.}~\bibnamefont
  {Wang}}, \bibinfo {author} {\bibfnamefont {J.}~\bibnamefont {Maisch}},
  \bibinfo {author} {\bibfnamefont {F.}~\bibnamefont {Tang}}, \bibinfo {author}
  {\bibfnamefont {D.}~\bibnamefont {Zhao}}, \bibinfo {author} {\bibfnamefont
  {S.}~\bibnamefont {Yang}}, \bibinfo {author} {\bibfnamefont {R.}~\bibnamefont
  {Joos}}, \bibinfo {author} {\bibfnamefont {S.~L.}\ \bibnamefont {Portalupi}},
  \bibinfo {author} {\bibfnamefont {P.}~\bibnamefont {Michler}},\ and\ \bibinfo
  {author} {\bibfnamefont {J.~H.}\ \bibnamefont {Smet}},\ }\bibfield  {title}
  {\bibinfo {title} {Highly {Polarized} {Single} {Photons} from
  {Strain}-{Induced} {Quasi}-{1D} {Localized} {Excitons} in
  {WSe}$_{\textrm{2}}$},\ }\href {https://doi.org/10.1021/acs.nanolett.1c01927}
  {\bibfield  {journal} {\bibinfo  {journal} {Nano Lett.}\ }\textbf {\bibinfo
  {volume} {21}},\ \bibinfo {pages} {7175} (\bibinfo {year}
  {2021})}\BibitemShut {NoStop}%
\bibitem [{\citenamefont {Paralikis}\ \emph {et~al.}(2024)\citenamefont
  {Paralikis}, \citenamefont {Piccinini}, \citenamefont {Madigawa},
  \citenamefont {Metuh}, \citenamefont {Vannucci}, \citenamefont {Gregersen},\
  and\ \citenamefont {Munkhbat}}]{paralikis_tailoring_2024}%
  \BibitemOpen
  \bibfield  {author} {\bibinfo {author} {\bibfnamefont {A.}~\bibnamefont
  {Paralikis}}, \bibinfo {author} {\bibfnamefont {C.}~\bibnamefont
  {Piccinini}}, \bibinfo {author} {\bibfnamefont {A.~A.}\ \bibnamefont
  {Madigawa}}, \bibinfo {author} {\bibfnamefont {P.}~\bibnamefont {Metuh}},
  \bibinfo {author} {\bibfnamefont {L.}~\bibnamefont {Vannucci}}, \bibinfo
  {author} {\bibfnamefont {N.}~\bibnamefont {Gregersen}},\ and\ \bibinfo
  {author} {\bibfnamefont {B.}~\bibnamefont {Munkhbat}},\ }\bibfield  {title}
  {\bibinfo {title} {Tailoring polarization in {WSe2} quantum emitters through
  deterministic strain engineering},\ }\href
  {https://doi.org/10.1038/s41699-024-00497-2} {\bibfield  {journal} {\bibinfo
  {journal} {npj 2D Mater Appl}\ }\textbf {\bibinfo {volume} {8}},\ \bibinfo
  {pages} {1} (\bibinfo {year} {2024})},\ \bibinfo {note} {publisher: Nature
  Publishing Group}\BibitemShut {NoStop}%
\bibitem [{\citenamefont {Tripathi}\ \emph {et~al.}(2018)\citenamefont
  {Tripathi}, \citenamefont {Iff}, \citenamefont {Betzold}, \citenamefont
  {Dusanowski}, \citenamefont {Emmerling}, \citenamefont {Moon}, \citenamefont
  {Lee}, \citenamefont {Kwon}, \citenamefont {Höfling},\ and\ \citenamefont
  {Schneider}}]{tripathi_spontaneous_2018}%
  \BibitemOpen
  \bibfield  {author} {\bibinfo {author} {\bibfnamefont {L.~N.}\ \bibnamefont
  {Tripathi}}, \bibinfo {author} {\bibfnamefont {O.}~\bibnamefont {Iff}},
  \bibinfo {author} {\bibfnamefont {S.}~\bibnamefont {Betzold}}, \bibinfo
  {author} {\bibfnamefont {L.}~\bibnamefont {Dusanowski}}, \bibinfo {author}
  {\bibfnamefont {M.}~\bibnamefont {Emmerling}}, \bibinfo {author}
  {\bibfnamefont {K.}~\bibnamefont {Moon}}, \bibinfo {author} {\bibfnamefont
  {Y.~J.}\ \bibnamefont {Lee}}, \bibinfo {author} {\bibfnamefont {S.-H.}\
  \bibnamefont {Kwon}}, \bibinfo {author} {\bibfnamefont {S.}~\bibnamefont
  {Höfling}},\ and\ \bibinfo {author} {\bibfnamefont {C.}~\bibnamefont
  {Schneider}},\ }\bibfield  {title} {\bibinfo {title} {Spontaneous {Emission}
  {Enhancement} in {Strain}-{Induced} {WSe}$_{\textrm{2}}$ {Monolayer}-{Based}
  {Quantum} {Light} {Sources} on {Metallic} {Surfaces}},\ }\href
  {https://doi.org/10.1021/acsphotonics.7b01053} {\bibfield  {journal}
  {\bibinfo  {journal} {ACS Photonics}\ }\textbf {\bibinfo {volume} {5}},\
  \bibinfo {pages} {1919} (\bibinfo {year} {2018})}\BibitemShut {NoStop}%
\bibitem [{\citenamefont {Shepard}\ \emph {et~al.}(2017)\citenamefont
  {Shepard}, \citenamefont {Ajayi}, \citenamefont {Li}, \citenamefont {Zhu},
  \citenamefont {Hone},\ and\ \citenamefont
  {Strauf}}]{shepard_nanobubble_2017}%
  \BibitemOpen
  \bibfield  {author} {\bibinfo {author} {\bibfnamefont {G.~D.}\ \bibnamefont
  {Shepard}}, \bibinfo {author} {\bibfnamefont {O.~A.}\ \bibnamefont {Ajayi}},
  \bibinfo {author} {\bibfnamefont {X.}~\bibnamefont {Li}}, \bibinfo {author}
  {\bibfnamefont {X.-Y.}\ \bibnamefont {Zhu}}, \bibinfo {author} {\bibfnamefont
  {J.}~\bibnamefont {Hone}},\ and\ \bibinfo {author} {\bibfnamefont
  {S.}~\bibnamefont {Strauf}},\ }\bibfield  {title} {\bibinfo {title}
  {Nanobubble induced formation of quantum emitters in monolayer
  semiconductors},\ }\href {https://doi.org/10.1088/2053-1583/aa629d}
  {\bibfield  {journal} {\bibinfo  {journal} {2D Mater.}\ }\textbf {\bibinfo
  {volume} {4}},\ \bibinfo {pages} {021019} (\bibinfo {year}
  {2017})}\BibitemShut {NoStop}%
\bibitem [{\citenamefont {Cianci}\ \emph {et~al.}(2023)\citenamefont {Cianci},
  \citenamefont {Blundo}, \citenamefont {Tuzi}, \citenamefont {Pettinari},
  \citenamefont {Olkowska-Pucko}, \citenamefont {Parmenopoulou}, \citenamefont
  {Peeters}, \citenamefont {Miriametro}, \citenamefont {Taniguchi},
  \citenamefont {Watanabe}, \citenamefont {Babinski}, \citenamefont {Molas},
  \citenamefont {Felici},\ and\ \citenamefont
  {Polimeni}}]{cianci_spatially_2023}%
  \BibitemOpen
  \bibfield  {author} {\bibinfo {author} {\bibfnamefont {S.}~\bibnamefont
  {Cianci}}, \bibinfo {author} {\bibfnamefont {E.}~\bibnamefont {Blundo}},
  \bibinfo {author} {\bibfnamefont {F.}~\bibnamefont {Tuzi}}, \bibinfo {author}
  {\bibfnamefont {G.}~\bibnamefont {Pettinari}}, \bibinfo {author}
  {\bibfnamefont {K.}~\bibnamefont {Olkowska-Pucko}}, \bibinfo {author}
  {\bibfnamefont {E.}~\bibnamefont {Parmenopoulou}}, \bibinfo {author}
  {\bibfnamefont {D.~B.~L.}\ \bibnamefont {Peeters}}, \bibinfo {author}
  {\bibfnamefont {A.}~\bibnamefont {Miriametro}}, \bibinfo {author}
  {\bibfnamefont {T.}~\bibnamefont {Taniguchi}}, \bibinfo {author}
  {\bibfnamefont {K.}~\bibnamefont {Watanabe}}, \bibinfo {author}
  {\bibfnamefont {A.}~\bibnamefont {Babinski}}, \bibinfo {author}
  {\bibfnamefont {M.~R.}\ \bibnamefont {Molas}}, \bibinfo {author}
  {\bibfnamefont {M.}~\bibnamefont {Felici}},\ and\ \bibinfo {author}
  {\bibfnamefont {A.}~\bibnamefont {Polimeni}},\ }\bibfield  {title} {\bibinfo
  {title} {Spatially {Controlled} {Single} {Photon} {Emitters} in
  {hBN}-{Capped} {WS2} {Domes}},\ }\href
  {https://doi.org/10.1002/adom.202202953} {\bibfield  {journal} {\bibinfo
  {journal} {Advanced Optical Materials}\ }\textbf {\bibinfo {volume} {11}},\
  \bibinfo {pages} {2202953} (\bibinfo {year} {2023})}\BibitemShut {NoStop}%
\bibitem [{\citenamefont {Darlington}\ \emph {et~al.}(2020)\citenamefont
  {Darlington}, \citenamefont {Carmesin}, \citenamefont {Florian},
  \citenamefont {Yanev}, \citenamefont {Ajayi}, \citenamefont {Ardelean},
  \citenamefont {Rhodes}, \citenamefont {Ghiotto}, \citenamefont {Krayev},
  \citenamefont {Watanabe}, \citenamefont {Taniguchi}, \citenamefont {Kysar},
  \citenamefont {Pasupathy}, \citenamefont {Hone}, \citenamefont {Jahnke},
  \citenamefont {Borys},\ and\ \citenamefont
  {Schuck}}]{darlington_imaging_2020}%
  \BibitemOpen
  \bibfield  {author} {\bibinfo {author} {\bibfnamefont {T.~P.}\ \bibnamefont
  {Darlington}}, \bibinfo {author} {\bibfnamefont {C.}~\bibnamefont
  {Carmesin}}, \bibinfo {author} {\bibfnamefont {M.}~\bibnamefont {Florian}},
  \bibinfo {author} {\bibfnamefont {E.}~\bibnamefont {Yanev}}, \bibinfo
  {author} {\bibfnamefont {O.}~\bibnamefont {Ajayi}}, \bibinfo {author}
  {\bibfnamefont {J.}~\bibnamefont {Ardelean}}, \bibinfo {author}
  {\bibfnamefont {D.~A.}\ \bibnamefont {Rhodes}}, \bibinfo {author}
  {\bibfnamefont {A.}~\bibnamefont {Ghiotto}}, \bibinfo {author} {\bibfnamefont
  {A.}~\bibnamefont {Krayev}}, \bibinfo {author} {\bibfnamefont
  {K.}~\bibnamefont {Watanabe}}, \bibinfo {author} {\bibfnamefont
  {T.}~\bibnamefont {Taniguchi}}, \bibinfo {author} {\bibfnamefont {J.~W.}\
  \bibnamefont {Kysar}}, \bibinfo {author} {\bibfnamefont {A.~N.}\ \bibnamefont
  {Pasupathy}}, \bibinfo {author} {\bibfnamefont {J.~C.}\ \bibnamefont {Hone}},
  \bibinfo {author} {\bibfnamefont {F.}~\bibnamefont {Jahnke}}, \bibinfo
  {author} {\bibfnamefont {N.~J.}\ \bibnamefont {Borys}},\ and\ \bibinfo
  {author} {\bibfnamefont {P.~J.}\ \bibnamefont {Schuck}},\ }\bibfield  {title}
  {\bibinfo {title} {Imaging strain-localized excitons in nanoscale bubbles of
  monolayer {WSe} 2 at room temperature},\ }\href
  {https://doi.org/10.1038/s41565-020-0730-5} {\bibfield  {journal} {\bibinfo
  {journal} {Nature Nanotechnology}\ }\textbf {\bibinfo {volume} {15}},\
  \bibinfo {pages} {854} (\bibinfo {year} {2020})},\ \bibinfo {note} {number:
  10 Publisher: Nature Publishing Group}\BibitemShut {NoStop}%
\bibitem [{\citenamefont {Preuss}\ \emph {et~al.}(2022)\citenamefont {Preuss},
  \citenamefont {Groll}, \citenamefont {Schmidt}, \citenamefont {Hahn},
  \citenamefont {Machnikowski}, \citenamefont {Bratschitsch}, \citenamefont
  {Kuhn}, \citenamefont {Vasconcellos},\ and\ \citenamefont
  {Wigger}}]{preuss_resonant_2022}%
  \BibitemOpen
  \bibfield  {author} {\bibinfo {author} {\bibfnamefont {J.~A.}\ \bibnamefont
  {Preuss}}, \bibinfo {author} {\bibfnamefont {D.}~\bibnamefont {Groll}},
  \bibinfo {author} {\bibfnamefont {R.}~\bibnamefont {Schmidt}}, \bibinfo
  {author} {\bibfnamefont {T.}~\bibnamefont {Hahn}}, \bibinfo {author}
  {\bibfnamefont {P.}~\bibnamefont {Machnikowski}}, \bibinfo {author}
  {\bibfnamefont {R.}~\bibnamefont {Bratschitsch}}, \bibinfo {author}
  {\bibfnamefont {T.}~\bibnamefont {Kuhn}}, \bibinfo {author} {\bibfnamefont
  {S.~M.~d.}\ \bibnamefont {Vasconcellos}},\ and\ \bibinfo {author}
  {\bibfnamefont {D.}~\bibnamefont {Wigger}},\ }\bibfield  {title} {\bibinfo
  {title} {Resonant and phonon-assisted ultrafast coherent control of a single
  {hBN} color center},\ }\href {https://doi.org/10.1364/OPTICA.448124}
  {\bibfield  {journal} {\bibinfo  {journal} {Optica}\ }\textbf {\bibinfo
  {volume} {9}},\ \bibinfo {pages} {522} (\bibinfo {year} {2022})},\ \bibinfo
  {note} {publisher: Optica Publishing Group}\BibitemShut {NoStop}%
\bibitem [{\citenamefont {Aharonovich}\ \emph {et~al.}(2022)\citenamefont
  {Aharonovich}, \citenamefont {Tetienne},\ and\ \citenamefont
  {Toth}}]{aharonovich_quantum_2022}%
  \BibitemOpen
  \bibfield  {author} {\bibinfo {author} {\bibfnamefont {I.}~\bibnamefont
  {Aharonovich}}, \bibinfo {author} {\bibfnamefont {J.-P.}\ \bibnamefont
  {Tetienne}},\ and\ \bibinfo {author} {\bibfnamefont {M.}~\bibnamefont
  {Toth}},\ }\bibfield  {title} {\bibinfo {title} {Quantum {Emitters} in
  {Hexagonal} {Boron} {Nitride}},\ }\href
  {https://doi.org/10.1021/acs.nanolett.2c03743} {\bibfield  {journal}
  {\bibinfo  {journal} {Nano Lett.}\ }\textbf {\bibinfo {volume} {22}},\
  \bibinfo {pages} {9227} (\bibinfo {year} {2022})}\BibitemShut {NoStop}%
\bibitem [{\citenamefont {Samaner}\ \emph {et~al.}(2022)\citenamefont
  {Samaner}, \citenamefont {Pacal}, \citenamefont {Mutlu}, \citenamefont
  {Uyanik},\ and\ \citenamefont {Ates}}]{samaner_freespace_2022}%
  \BibitemOpen
  \bibfield  {author} {\bibinfo {author} {\bibfnamefont {C.}~\bibnamefont
  {Samaner}}, \bibinfo {author} {\bibfnamefont {S.}~\bibnamefont {Pacal}},
  \bibinfo {author} {\bibfnamefont {G.}~\bibnamefont {Mutlu}}, \bibinfo
  {author} {\bibfnamefont {K.}~\bibnamefont {Uyanik}},\ and\ \bibinfo {author}
  {\bibfnamefont {S.}~\bibnamefont {Ates}},\ }\bibfield  {title} {\bibinfo
  {title} {Free‐{Space} {Quantum} {Key} {Distribution} with {Single}
  {Photons} from {Defects} in {Hexagonal} {Boron} {Nitride}},\ }\href
  {https://doi.org/10.1002/qute.202200059} {\bibfield  {journal} {\bibinfo
  {journal} {Adv Quantum Tech}\ }\textbf {\bibinfo {volume} {5}},\ \bibinfo
  {pages} {2200059} (\bibinfo {year} {2022})}\BibitemShut {NoStop}%
\bibitem [{\citenamefont {Kretzschmar}\ \emph {et~al.}(2024)\citenamefont
  {Kretzschmar}, \citenamefont {Ritter}, \citenamefont {Kumar}, \citenamefont
  {Vogl}, \citenamefont {Eilenberger},\ and\ \citenamefont
  {Schmidt}}]{kretzschmar_quantitative_2024}%
  \BibitemOpen
  \bibfield  {author} {\bibinfo {author} {\bibfnamefont {T.}~\bibnamefont
  {Kretzschmar}}, \bibinfo {author} {\bibfnamefont {S.}~\bibnamefont {Ritter}},
  \bibinfo {author} {\bibfnamefont {A.}~\bibnamefont {Kumar}}, \bibinfo
  {author} {\bibfnamefont {T.}~\bibnamefont {Vogl}}, \bibinfo {author}
  {\bibfnamefont {F.}~\bibnamefont {Eilenberger}},\ and\ \bibinfo {author}
  {\bibfnamefont {F.}~\bibnamefont {Schmidt}},\ }\bibfield  {title} {\bibinfo
  {title} {Quantitative {Investigation} of {Quantum} {Emitter} {Yield} in
  {Drop}-{Casted} {Hexagonal} {Boron} {Nitride} {Nanoflakes}},\ }\href
  {https://doi.org/10.1021/acsaom.4c00200} {\bibfield  {journal} {\bibinfo
  {journal} {ACS Appl. Opt. Mater.}\ }\textbf {\bibinfo {volume} {2}},\
  \bibinfo {pages} {1427} (\bibinfo {year} {2024})},\ \bibinfo {note}
  {publisher: American Chemical Society}\BibitemShut {NoStop}%
\bibitem [{\citenamefont {Drawer}\ \emph {et~al.}(2023)\citenamefont {Drawer},
  \citenamefont {Mitryakhin}, \citenamefont {Shan}, \citenamefont {Stephan},
  \citenamefont {Gittinger}, \citenamefont {Lackner}, \citenamefont {Han},
  \citenamefont {Leibeling}, \citenamefont {Eilenberger}, \citenamefont
  {Banerjee}, \citenamefont {Tongay}, \citenamefont {Watanabe}, \citenamefont
  {Taniguchi}, \citenamefont {Lienau}, \citenamefont {Silies}, \citenamefont
  {Anton-Solanas}, \citenamefont {Esmann},\ and\ \citenamefont
  {Schneider}}]{drawer_monolayer-based_2023}%
  \BibitemOpen
  \bibfield  {author} {\bibinfo {author} {\bibfnamefont {J.-C.}\ \bibnamefont
  {Drawer}}, \bibinfo {author} {\bibfnamefont {V.~N.}\ \bibnamefont
  {Mitryakhin}}, \bibinfo {author} {\bibfnamefont {H.}~\bibnamefont {Shan}},
  \bibinfo {author} {\bibfnamefont {S.}~\bibnamefont {Stephan}}, \bibinfo
  {author} {\bibfnamefont {M.}~\bibnamefont {Gittinger}}, \bibinfo {author}
  {\bibfnamefont {L.}~\bibnamefont {Lackner}}, \bibinfo {author} {\bibfnamefont
  {B.}~\bibnamefont {Han}}, \bibinfo {author} {\bibfnamefont {G.}~\bibnamefont
  {Leibeling}}, \bibinfo {author} {\bibfnamefont {F.}~\bibnamefont
  {Eilenberger}}, \bibinfo {author} {\bibfnamefont {R.}~\bibnamefont
  {Banerjee}}, \bibinfo {author} {\bibfnamefont {S.}~\bibnamefont {Tongay}},
  \bibinfo {author} {\bibfnamefont {K.}~\bibnamefont {Watanabe}}, \bibinfo
  {author} {\bibfnamefont {T.}~\bibnamefont {Taniguchi}}, \bibinfo {author}
  {\bibfnamefont {C.}~\bibnamefont {Lienau}}, \bibinfo {author} {\bibfnamefont
  {M.}~\bibnamefont {Silies}}, \bibinfo {author} {\bibfnamefont
  {C.}~\bibnamefont {Anton-Solanas}}, \bibinfo {author} {\bibfnamefont
  {M.}~\bibnamefont {Esmann}},\ and\ \bibinfo {author} {\bibfnamefont
  {C.}~\bibnamefont {Schneider}},\ }\bibfield  {title} {\bibinfo {title}
  {Monolayer-based single-photon source in a liquid-helium-free open cavity
  featuring 65\% brightness and quantum coherence},\ }\href
  {https://doi.org/10.1021/acs.nanolett.3c02584} {\bibfield  {journal}
  {\bibinfo  {journal} {Nano Letters}\ }\textbf {\bibinfo {volume} {23}},\
  \bibinfo {pages} {8683} (\bibinfo {year} {2023})}\BibitemShut {NoStop}%
\bibitem [{\citenamefont {Fournier}\ \emph {et~al.}(2023)\citenamefont
  {Fournier}, \citenamefont {Roux}, \citenamefont {Watanabe}, \citenamefont
  {Taniguchi}, \citenamefont {Buil}, \citenamefont {Barjon}, \citenamefont
  {Hermier},\ and\ \citenamefont {Delteil}}]{fournier_two-photon_2023}%
  \BibitemOpen
  \bibfield  {author} {\bibinfo {author} {\bibfnamefont {C.}~\bibnamefont
  {Fournier}}, \bibinfo {author} {\bibfnamefont {S.}~\bibnamefont {Roux}},
  \bibinfo {author} {\bibfnamefont {K.}~\bibnamefont {Watanabe}}, \bibinfo
  {author} {\bibfnamefont {T.}~\bibnamefont {Taniguchi}}, \bibinfo {author}
  {\bibfnamefont {S.}~\bibnamefont {Buil}}, \bibinfo {author} {\bibfnamefont
  {J.}~\bibnamefont {Barjon}}, \bibinfo {author} {\bibfnamefont {J.-P.}\
  \bibnamefont {Hermier}},\ and\ \bibinfo {author} {\bibfnamefont
  {A.}~\bibnamefont {Delteil}},\ }\bibfield  {title} {\bibinfo {title}
  {Two-photon interference from a quantum emitter in hexagonal boron nitride},\
  }\href {https://doi.org/10.1103/PhysRevApplied.19.L041003} {\bibfield
  {journal} {\bibinfo  {journal} {Physical Review Applied}\ }\textbf {\bibinfo
  {volume} {19}},\ \bibinfo {pages} {L041003} (\bibinfo {year} {2023})},\
  \bibinfo {note} {publisher: American Physical Society}\BibitemShut {NoStop}%
\bibitem [{\citenamefont {Vannucci}\ \emph {et~al.}(2024)\citenamefont
  {Vannucci}, \citenamefont {Neto}, \citenamefont {Piccinini}, \citenamefont
  {Paralikis}, \citenamefont {Gregersen},\ and\ \citenamefont
  {Munkhbat}}]{vannucci_single-photon_2024}%
  \BibitemOpen
  \bibfield  {author} {\bibinfo {author} {\bibfnamefont {L.}~\bibnamefont
  {Vannucci}}, \bibinfo {author} {\bibfnamefont {J.~F.}\ \bibnamefont {Neto}},
  \bibinfo {author} {\bibfnamefont {C.}~\bibnamefont {Piccinini}}, \bibinfo
  {author} {\bibfnamefont {A.}~\bibnamefont {Paralikis}}, \bibinfo {author}
  {\bibfnamefont {N.}~\bibnamefont {Gregersen}},\ and\ \bibinfo {author}
  {\bibfnamefont {B.}~\bibnamefont {Munkhbat}},\ }\bibfield  {title} {\bibinfo
  {title} {Single-photon emitters in {${\mathrm{WSe}}_{2}$:} critical role of
  phonons on excitation schemes and indistinguishability},\ }\href
  {https://doi.org/10.1103/PhysRevB.109.245304} {\bibfield  {journal} {\bibinfo
   {journal} {Physical Review B}\ }\textbf {\bibinfo {volume} {109}},\ \bibinfo
  {pages} {245304} (\bibinfo {year} {2024})},\ \bibinfo {note} {publisher:
  American Physical Society}\BibitemShut {NoStop}%
\bibitem [{\citenamefont {Mitryakhin}\ \emph {et~al.}(2024)\citenamefont
  {Mitryakhin}, \citenamefont {Steinhoff}, \citenamefont {Drawer},
  \citenamefont {Shan}, \citenamefont {Florian}, \citenamefont {Lackner},
  \citenamefont {Han}, \citenamefont {Eilenberger}, \citenamefont {Tongay},
  \citenamefont {Watanabe}, \citenamefont {Taniguchi}, \citenamefont
  {Antón-Solanas}, \citenamefont {Predojević}, \citenamefont {Gies},
  \citenamefont {Esmann},\ and\ \citenamefont
  {Schneider}}]{mitryakhin_engineering_2024}%
  \BibitemOpen
  \bibfield  {author} {\bibinfo {author} {\bibfnamefont {V.~N.}\ \bibnamefont
  {Mitryakhin}}, \bibinfo {author} {\bibfnamefont {A.}~\bibnamefont
  {Steinhoff}}, \bibinfo {author} {\bibfnamefont {J.-C.}\ \bibnamefont
  {Drawer}}, \bibinfo {author} {\bibfnamefont {H.}~\bibnamefont {Shan}},
  \bibinfo {author} {\bibfnamefont {M.}~\bibnamefont {Florian}}, \bibinfo
  {author} {\bibfnamefont {L.}~\bibnamefont {Lackner}}, \bibinfo {author}
  {\bibfnamefont {B.}~\bibnamefont {Han}}, \bibinfo {author} {\bibfnamefont
  {F.}~\bibnamefont {Eilenberger}}, \bibinfo {author} {\bibfnamefont {S.~A.}\
  \bibnamefont {Tongay}}, \bibinfo {author} {\bibfnamefont {K.}~\bibnamefont
  {Watanabe}}, \bibinfo {author} {\bibfnamefont {T.}~\bibnamefont {Taniguchi}},
  \bibinfo {author} {\bibfnamefont {C.}~\bibnamefont {Antón-Solanas}},
  \bibinfo {author} {\bibfnamefont {A.}~\bibnamefont {Predojević}}, \bibinfo
  {author} {\bibfnamefont {C.}~\bibnamefont {Gies}}, \bibinfo {author}
  {\bibfnamefont {M.}~\bibnamefont {Esmann}},\ and\ \bibinfo {author}
  {\bibfnamefont {C.}~\bibnamefont {Schneider}},\ }\bibfield  {title} {\bibinfo
  {title} {Engineering the impact of phonon dephasing on the coherence of a
  {${\mathrm{WSe}}_{2}$} single-photon source via cavity quantum
  electrodynamics},\ }\href {https://doi.org/10.1103/PhysRevLett.132.206903}
  {\bibfield  {journal} {\bibinfo  {journal} {Physical Review Letters}\
  }\textbf {\bibinfo {volume} {132}},\ \bibinfo {pages} {206903} (\bibinfo
  {year} {2024})},\ \bibinfo {note} {publisher: American Physical
  Society}\BibitemShut {NoStop}%
\bibitem [{\citenamefont {Khatri}\ \emph {et~al.}(2019)\citenamefont {Khatri},
  \citenamefont {Luxmoore},\ and\ \citenamefont {Ramsay}}]{khatri_phonon_2019}%
  \BibitemOpen
  \bibfield  {author} {\bibinfo {author} {\bibfnamefont {P.}~\bibnamefont
  {Khatri}}, \bibinfo {author} {\bibfnamefont {I.~J.}\ \bibnamefont
  {Luxmoore}},\ and\ \bibinfo {author} {\bibfnamefont {A.~J.}\ \bibnamefont
  {Ramsay}},\ }\bibfield  {title} {\bibinfo {title} {Phonon sidebands of color
  centers in hexagonal boron nitride},\ }\href
  {https://doi.org/10.1103/PhysRevB.100.125305} {\bibfield  {journal} {\bibinfo
   {journal} {Physical Review B}\ }\textbf {\bibinfo {volume} {100}},\ \bibinfo
  {pages} {125305} (\bibinfo {year} {2019})},\ \bibinfo {note} {publisher:
  American Physical Society}\BibitemShut {NoStop}%
\bibitem [{\citenamefont {Denning}\ \emph {et~al.}(2022)\citenamefont
  {Denning}, \citenamefont {Wubs}, \citenamefont {Stenger}, \citenamefont
  {Mørk},\ and\ \citenamefont {Kristensen}}]{denning_quantum_2022}%
  \BibitemOpen
  \bibfield  {author} {\bibinfo {author} {\bibfnamefont {E.~V.}\ \bibnamefont
  {Denning}}, \bibinfo {author} {\bibfnamefont {M.}~\bibnamefont {Wubs}},
  \bibinfo {author} {\bibfnamefont {N.}~\bibnamefont {Stenger}}, \bibinfo
  {author} {\bibfnamefont {J.}~\bibnamefont {Mørk}},\ and\ \bibinfo {author}
  {\bibfnamefont {P.~T.}\ \bibnamefont {Kristensen}},\ }\bibfield  {title}
  {\bibinfo {title} {Quantum theory of two-dimensional materials coupled to
  electromagnetic resonators},\ }\href
  {https://doi.org/10.1103/PhysRevB.105.085306} {\bibfield  {journal} {\bibinfo
   {journal} {Physical Review B}\ }\textbf {\bibinfo {volume} {105}},\ \bibinfo
  {pages} {085306} (\bibinfo {year} {2022})},\ \bibinfo {note} {publisher:
  American Physical Society}\BibitemShut {NoStop}%
\bibitem [{\citenamefont {Svendsen}\ \emph {et~al.}(2023)\citenamefont
  {Svendsen}, \citenamefont {Ali}, \citenamefont {Stenger}, \citenamefont
  {Thygesen},\ and\ \citenamefont {Iles-Smith}}]{svendsen_signatures_2023}%
  \BibitemOpen
  \bibfield  {author} {\bibinfo {author} {\bibfnamefont {M.~K.}\ \bibnamefont
  {Svendsen}}, \bibinfo {author} {\bibfnamefont {S.}~\bibnamefont {Ali}},
  \bibinfo {author} {\bibfnamefont {N.}~\bibnamefont {Stenger}}, \bibinfo
  {author} {\bibfnamefont {K.~S.}\ \bibnamefont {Thygesen}},\ and\ \bibinfo
  {author} {\bibfnamefont {J.}~\bibnamefont {Iles-Smith}},\ }\bibfield  {title}
  {\bibinfo {title} {Signatures of non-markovianity in cavity {QED} with color
  centers in two-dimensional materials},\ }\href
  {https://doi.org/10.1103/PhysRevResearch.5.L032037} {\bibfield  {journal}
  {\bibinfo  {journal} {Physical Review Research}\ }\textbf {\bibinfo {volume}
  {5}},\ \bibinfo {pages} {L032037} (\bibinfo {year} {2023})},\ \bibinfo {note}
  {publisher: American Physical Society}\BibitemShut {NoStop}%
\bibitem [{\citenamefont {Piccinini}\ \emph {et~al.}(2024)\citenamefont
  {Piccinini}, \citenamefont {Paralikis}, \citenamefont {Neto}, \citenamefont
  {Madigawa}, \citenamefont {Wyborski}, \citenamefont {Remesh}, \citenamefont
  {Vannucci}, \citenamefont {Gregersen},\ and\ \citenamefont
  {Munkhbat}}]{piccinini_high-purity_2024}%
  \BibitemOpen
  \bibfield  {author} {\bibinfo {author} {\bibfnamefont {C.}~\bibnamefont
  {Piccinini}}, \bibinfo {author} {\bibfnamefont {A.}~\bibnamefont
  {Paralikis}}, \bibinfo {author} {\bibfnamefont {J.~F.}\ \bibnamefont {Neto}},
  \bibinfo {author} {\bibfnamefont {A.~A.}\ \bibnamefont {Madigawa}}, \bibinfo
  {author} {\bibfnamefont {P.}~\bibnamefont {Wyborski}}, \bibinfo {author}
  {\bibfnamefont {V.}~\bibnamefont {Remesh}}, \bibinfo {author} {\bibfnamefont
  {L.}~\bibnamefont {Vannucci}}, \bibinfo {author} {\bibfnamefont
  {N.}~\bibnamefont {Gregersen}},\ and\ \bibinfo {author} {\bibfnamefont
  {B.}~\bibnamefont {Munkhbat}},\ }\href
  {https://doi.org/10.48550/arXiv.2406.07097} {\bibinfo {title} {High-purity
  and stable single-photon emission in bilayer {WSe$_2$} via phonon-assisted
  excitation}} (\bibinfo {year} {2024}),\ \bibinfo {note}
  {{arXiv:2406.07097}}\BibitemShut {NoStop}%
\bibitem [{\citenamefont {Grange}\ \emph {et~al.}(2015)\citenamefont {Grange},
  \citenamefont {Hornecker}, \citenamefont {Hunger}, \citenamefont {Poizat},
  \citenamefont {Gérard}, \citenamefont {Senellart},\ and\ \citenamefont
  {Auffèves}}]{grange_cavity-funneled_2015}%
  \BibitemOpen
  \bibfield  {author} {\bibinfo {author} {\bibfnamefont {T.}~\bibnamefont
  {Grange}}, \bibinfo {author} {\bibfnamefont {G.}~\bibnamefont {Hornecker}},
  \bibinfo {author} {\bibfnamefont {D.}~\bibnamefont {Hunger}}, \bibinfo
  {author} {\bibfnamefont {J.-P.}\ \bibnamefont {Poizat}}, \bibinfo {author}
  {\bibfnamefont {J.-M.}\ \bibnamefont {Gérard}}, \bibinfo {author}
  {\bibfnamefont {P.}~\bibnamefont {Senellart}},\ and\ \bibinfo {author}
  {\bibfnamefont {A.}~\bibnamefont {Auffèves}},\ }\bibfield  {title} {\bibinfo
  {title} {Cavity-funneled generation of indistinguishable single photons from
  strongly dissipative quantum emitters},\ }\href
  {https://doi.org/10.1103/PhysRevLett.114.193601} {\bibfield  {journal}
  {\bibinfo  {journal} {Physical Review Letters}\ }\textbf {\bibinfo {volume}
  {114}},\ \bibinfo {pages} {193601} (\bibinfo {year} {2015})},\ \bibinfo
  {note} {publisher: American Physical Society}\BibitemShut {NoStop}%
\bibitem [{\citenamefont {Krummheuer}\ \emph {et~al.}(2002)\citenamefont
  {Krummheuer}, \citenamefont {Axt},\ and\ \citenamefont
  {Kuhn}}]{krummheuer_theory_2002}%
  \BibitemOpen
  \bibfield  {author} {\bibinfo {author} {\bibfnamefont {B.}~\bibnamefont
  {Krummheuer}}, \bibinfo {author} {\bibfnamefont {V.~M.}\ \bibnamefont
  {Axt}},\ and\ \bibinfo {author} {\bibfnamefont {T.}~\bibnamefont {Kuhn}},\
  }\bibfield  {title} {\bibinfo {title} {Theory of pure dephasing and the
  resulting absorption line shape in semiconductor quantum dots},\ }\href
  {https://doi.org/10.1103/PhysRevB.65.195313} {\bibfield  {journal} {\bibinfo
  {journal} {Physical Review B}\ }\textbf {\bibinfo {volume} {65}},\ \bibinfo
  {pages} {195313} (\bibinfo {year} {2002})}\BibitemShut {NoStop}%
\bibitem [{\citenamefont {Ferreira~Neto}\ \emph {et~al.}(2024)\citenamefont
  {Ferreira~Neto}, \citenamefont {Bundgaard-Nielsen}, \citenamefont
  {Gregersen},\ and\ \citenamefont
  {Vannucci}}]{ferreira_neto_one-dimensional_2024}%
  \BibitemOpen
  \bibfield  {author} {\bibinfo {author} {\bibfnamefont {J.}~\bibnamefont
  {Ferreira~Neto}}, \bibinfo {author} {\bibfnamefont {M.}~\bibnamefont
  {Bundgaard-Nielsen}}, \bibinfo {author} {\bibfnamefont {N.}~\bibnamefont
  {Gregersen}},\ and\ \bibinfo {author} {\bibfnamefont {L.}~\bibnamefont
  {Vannucci}},\ }\bibfield  {title} {\bibinfo {title} {One-dimensional photonic
  wire as a single-photon source: Implications of cavity {QED} to a phonon bath
  of reduced dimensionality},\ }\href
  {https://doi.org/10.1103/PhysRevB.110.115308} {\bibfield  {journal} {\bibinfo
   {journal} {Physical Review B}\ }\textbf {\bibinfo {volume} {110}},\ \bibinfo
  {pages} {115308} (\bibinfo {year} {2024})},\ \bibinfo {note} {publisher:
  American Physical Society}\BibitemShut {NoStop}%
\bibitem [{\citenamefont {Kaer}\ \emph {et~al.}(2013)\citenamefont {Kaer},
  \citenamefont {Lodahl}, \citenamefont {Jauho},\ and\ \citenamefont
  {Mork}}]{kaer_microscopic_2013}%
  \BibitemOpen
  \bibfield  {author} {\bibinfo {author} {\bibfnamefont {P.}~\bibnamefont
  {Kaer}}, \bibinfo {author} {\bibfnamefont {P.}~\bibnamefont {Lodahl}},
  \bibinfo {author} {\bibfnamefont {A.-P.}\ \bibnamefont {Jauho}},\ and\
  \bibinfo {author} {\bibfnamefont {J.}~\bibnamefont {Mork}},\ }\bibfield
  {title} {\bibinfo {title} {Microscopic theory of indistinguishable
  single-photon emission from a quantum dot coupled to a cavity: The role of
  non-markovian phonon-induced decoherence},\ }\href
  {https://doi.org/10.1103/PhysRevB.87.081308} {\bibfield  {journal} {\bibinfo
  {journal} {Physical Review B}\ }\textbf {\bibinfo {volume} {87}},\ \bibinfo
  {pages} {081308} (\bibinfo {year} {2013})},\ \bibinfo {note} {publisher:
  American Physical Society}\BibitemShut {NoStop}%
\bibitem [{\citenamefont {Hohenester}(2007)}]{hohenester_quantum_2007}%
  \BibitemOpen
  \bibfield  {author} {\bibinfo {author} {\bibfnamefont {U.}~\bibnamefont
  {Hohenester}},\ }\bibfield  {title} {\bibinfo {title} {Quantum control of
  polaron states in semiconductor quantum dots},\ }\href
  {https://doi.org/10.1088/0953-4075/40/11/S06} {\bibfield  {journal} {\bibinfo
   {journal} {Journal of Physics B: Atomic, Molecular and Optical Physics}\
  }\textbf {\bibinfo {volume} {40}},\ \bibinfo {pages} {S315} (\bibinfo {year}
  {2007})}\BibitemShut {NoStop}%
\bibitem [{\citenamefont {Zimmermann}\ and\ \citenamefont
  {Runge}(2002)}]{zimmermann_dephasing_2002}%
  \BibitemOpen
  \bibfield  {author} {\bibinfo {author} {\bibfnamefont {R.}~\bibnamefont
  {Zimmermann}}\ and\ \bibinfo {author} {\bibfnamefont {E.}~\bibnamefont
  {Runge}},\ }\bibfield  {title} {\bibinfo {title} {Dephasing in quantum dots
  via electron-phonon interaction},\ }\href@noop {} {\bibfield  {journal}
  {\bibinfo  {journal} {Proc. 26th {ICPS}, Edinburgh}\ } (\bibinfo {year}
  {2002})}\BibitemShut {NoStop}%
\bibitem [{\citenamefont {Flatten}\ \emph {et~al.}(2018)\citenamefont
  {Flatten}, \citenamefont {Weng}, \citenamefont {Branny}, \citenamefont
  {Johnson}, \citenamefont {Dolan}, \citenamefont {Trichet}, \citenamefont
  {Gerardot},\ and\ \citenamefont {Smith}}]{Flatten2018}%
  \BibitemOpen
  \bibfield  {author} {\bibinfo {author} {\bibfnamefont {L.~C.}\ \bibnamefont
  {Flatten}}, \bibinfo {author} {\bibfnamefont {L.}~\bibnamefont {Weng}},
  \bibinfo {author} {\bibfnamefont {A.}~\bibnamefont {Branny}}, \bibinfo
  {author} {\bibfnamefont {S.}~\bibnamefont {Johnson}}, \bibinfo {author}
  {\bibfnamefont {P.~R.}\ \bibnamefont {Dolan}}, \bibinfo {author}
  {\bibfnamefont {A.~A.}\ \bibnamefont {Trichet}}, \bibinfo {author}
  {\bibfnamefont {B.~D.}\ \bibnamefont {Gerardot}},\ and\ \bibinfo {author}
  {\bibfnamefont {J.~M.}\ \bibnamefont {Smith}},\ }\bibfield  {title} {\bibinfo
  {title} {{Microcavity enhanced single photon emission from two-dimensional
  WSe2}},\ }\bibfield  {journal} {\bibinfo  {journal} {Appl. Phys. Lett.}\
  }\textbf {\bibinfo {volume} {112}},\ \href
  {https://doi.org/10.1063/1.5026779} {10.1063/1.5026779} (\bibinfo {year}
  {2018}),\ \Eprint {https://arxiv.org/abs/1807.02778} {arXiv:1807.02778}
  \BibitemShut {NoStop}%
\bibitem [{\citenamefont {Iff}\ \emph {et~al.}(2021)\citenamefont {Iff},
  \citenamefont {Buchinger}, \citenamefont {Mocza{\l}a-Dusanowska},
  \citenamefont {Kamp}, \citenamefont {Betzold}, \citenamefont {Davanco},
  \citenamefont {Srinivasan}, \citenamefont {Tongay}, \citenamefont
  {Ant{\'{o}}n-Solanas}, \citenamefont {H{\"{o}}fling},\ and\ \citenamefont
  {Schneider}}]{Iff2021}%
  \BibitemOpen
  \bibfield  {author} {\bibinfo {author} {\bibfnamefont {O.}~\bibnamefont
  {Iff}}, \bibinfo {author} {\bibfnamefont {Q.}~\bibnamefont {Buchinger}},
  \bibinfo {author} {\bibfnamefont {M.}~\bibnamefont {Mocza{\l}a-Dusanowska}},
  \bibinfo {author} {\bibfnamefont {M.}~\bibnamefont {Kamp}}, \bibinfo {author}
  {\bibfnamefont {S.}~\bibnamefont {Betzold}}, \bibinfo {author} {\bibfnamefont
  {M.}~\bibnamefont {Davanco}}, \bibinfo {author} {\bibfnamefont
  {K.}~\bibnamefont {Srinivasan}}, \bibinfo {author} {\bibfnamefont
  {S.}~\bibnamefont {Tongay}}, \bibinfo {author} {\bibfnamefont
  {C.}~\bibnamefont {Ant{\'{o}}n-Solanas}}, \bibinfo {author} {\bibfnamefont
  {S.}~\bibnamefont {H{\"{o}}fling}},\ and\ \bibinfo {author} {\bibfnamefont
  {C.}~\bibnamefont {Schneider}},\ }\bibfield  {title} {\bibinfo {title}
  {{Purcell-Enhanced Single Photon Source Based on a Deterministically Placed
  WSe2Monolayer Quantum Dot in a Circular Bragg Grating Cavity}},\ }\href
  {https://doi.org/10.1021/acs.nanolett.1c00978} {\bibfield  {journal}
  {\bibinfo  {journal} {Nano Lett.}\ }\textbf {\bibinfo {volume} {21}},\
  \bibinfo {pages} {4715} (\bibinfo {year} {2021})},\ \Eprint
  {https://arxiv.org/abs/2102.02827} {arXiv:2102.02827} \BibitemShut {NoStop}%
\bibitem [{\citenamefont {Tran}\ \emph {et~al.}(2017)\citenamefont {Tran},
  \citenamefont {Wang}, \citenamefont {Xu}, \citenamefont {Yang}, \citenamefont
  {Toth}, \citenamefont {Odom},\ and\ \citenamefont {Aharonovich}}]{Tran2017}%
  \BibitemOpen
  \bibfield  {author} {\bibinfo {author} {\bibfnamefont {T.~T.}\ \bibnamefont
  {Tran}}, \bibinfo {author} {\bibfnamefont {D.}~\bibnamefont {Wang}}, \bibinfo
  {author} {\bibfnamefont {Z.~Q.}\ \bibnamefont {Xu}}, \bibinfo {author}
  {\bibfnamefont {A.}~\bibnamefont {Yang}}, \bibinfo {author} {\bibfnamefont
  {M.}~\bibnamefont {Toth}}, \bibinfo {author} {\bibfnamefont {T.~W.}\
  \bibnamefont {Odom}},\ and\ \bibinfo {author} {\bibfnamefont
  {I.}~\bibnamefont {Aharonovich}},\ }\bibfield  {title} {\bibinfo {title}
  {{Deterministic Coupling of Quantum Emitters in 2D Materials to Plasmonic
  Nanocavity Arrays}},\ }\href {https://doi.org/10.1021/acs.nanolett.7b00444}
  {\bibfield  {journal} {\bibinfo  {journal} {Nano Lett.}\ }\textbf {\bibinfo
  {volume} {17}},\ \bibinfo {pages} {2634} (\bibinfo {year}
  {2017})}\BibitemShut {NoStop}%
\bibitem [{\citenamefont {Sakib}\ \emph {et~al.}(2024)\citenamefont {Sakib},
  \citenamefont {Triplett}, \citenamefont {Harris}, \citenamefont {Hussain},
  \citenamefont {Senichev}, \citenamefont {Momenzadeh}, \citenamefont
  {Bocanegra}, \citenamefont {Vabishchevich}, \citenamefont {Wu}, \citenamefont
  {Boltasseva}, \citenamefont {Shalaev},\ and\ \citenamefont
  {Shcherbakov}}]{Sakib2024}%
  \BibitemOpen
  \bibfield  {author} {\bibinfo {author} {\bibfnamefont {M.~A.}\ \bibnamefont
  {Sakib}}, \bibinfo {author} {\bibfnamefont {B.}~\bibnamefont {Triplett}},
  \bibinfo {author} {\bibfnamefont {W.}~\bibnamefont {Harris}}, \bibinfo
  {author} {\bibfnamefont {N.}~\bibnamefont {Hussain}}, \bibinfo {author}
  {\bibfnamefont {A.}~\bibnamefont {Senichev}}, \bibinfo {author}
  {\bibfnamefont {M.}~\bibnamefont {Momenzadeh}}, \bibinfo {author}
  {\bibfnamefont {J.}~\bibnamefont {Bocanegra}}, \bibinfo {author}
  {\bibfnamefont {P.}~\bibnamefont {Vabishchevich}}, \bibinfo {author}
  {\bibfnamefont {R.}~\bibnamefont {Wu}}, \bibinfo {author} {\bibfnamefont
  {A.}~\bibnamefont {Boltasseva}}, \bibinfo {author} {\bibfnamefont {V.~M.}\
  \bibnamefont {Shalaev}},\ and\ \bibinfo {author} {\bibfnamefont {M.~R.}\
  \bibnamefont {Shcherbakov}},\ }\bibfield  {title} {\bibinfo {title}
  {{Purcell-Induced Bright Single Photon Emitters in Hexagonal Boron
  Nitride}},\ }\href {https://doi.org/10.1021/acs.nanolett.4c02581} {\bibfield
  {journal} {\bibinfo  {journal} {Nano Lett.}\ }\textbf {\bibinfo {volume}
  {24}},\ \bibinfo {pages} {12390} (\bibinfo {year} {2024})},\ \Eprint
  {https://arxiv.org/abs/2405.02516} {arXiv:2405.02516} \BibitemShut {NoStop}%
\bibitem [{\citenamefont {Luo}\ \emph {et~al.}(2018)\citenamefont {Luo},
  \citenamefont {Shepard}, \citenamefont {Ardelean}, \citenamefont {Rhodes},
  \citenamefont {Kim}, \citenamefont {Barmak}, \citenamefont {Hone},\ and\
  \citenamefont {Strauf}}]{Luo2018}%
  \BibitemOpen
  \bibfield  {author} {\bibinfo {author} {\bibfnamefont {Y.}~\bibnamefont
  {Luo}}, \bibinfo {author} {\bibfnamefont {G.~D.}\ \bibnamefont {Shepard}},
  \bibinfo {author} {\bibfnamefont {J.~V.}\ \bibnamefont {Ardelean}}, \bibinfo
  {author} {\bibfnamefont {D.~A.}\ \bibnamefont {Rhodes}}, \bibinfo {author}
  {\bibfnamefont {B.}~\bibnamefont {Kim}}, \bibinfo {author} {\bibfnamefont
  {K.}~\bibnamefont {Barmak}}, \bibinfo {author} {\bibfnamefont {J.~C.}\
  \bibnamefont {Hone}},\ and\ \bibinfo {author} {\bibfnamefont
  {S.}~\bibnamefont {Strauf}},\ }\bibfield  {title} {\bibinfo {title}
  {{Deterministic coupling of site-controlled quantum emitters in monolayer
  WSe2 to plasmonic nanocavities}},\ }\href
  {https://doi.org/10.1038/s41565-018-0275-z} {\bibfield  {journal} {\bibinfo
  {journal} {Nat. Nanotechnol.}\ }\textbf {\bibinfo {volume} {13}},\ \bibinfo
  {pages} {1137} (\bibinfo {year} {2018})}\BibitemShut {NoStop}%
\bibitem [{\citenamefont {Lindblad}(1976)}]{lindblad_generators_1976}%
  \BibitemOpen
  \bibfield  {author} {\bibinfo {author} {\bibfnamefont {G.}~\bibnamefont
  {Lindblad}},\ }\bibfield  {title} {\bibinfo {title} {On the generators of
  quantum dynamical semigroups},\ }\href {https://doi.org/10.1007/BF01608499}
  {\bibfield  {journal} {\bibinfo  {journal} {Communications in Mathematical
  Physics}\ }\textbf {\bibinfo {volume} {48}},\ \bibinfo {pages} {119}
  (\bibinfo {year} {1976})}\BibitemShut {NoStop}%
\bibitem [{\citenamefont {Breuer}(2007)}]{breuer_theory_2007}%
  \BibitemOpen
  \bibfield  {author} {\bibinfo {author} {\bibfnamefont {H.-P.}\ \bibnamefont
  {Breuer}},\ }\href@noop {} {\emph {\bibinfo {title} {The Theory of Open
  Quantum Systems}}}\ (\bibinfo  {publisher} {Oxford University Press, {USA}},\
  \bibinfo {year} {2007})\BibitemShut {NoStop}%
\bibitem [{\citenamefont {Carmichael}(1999)}]{carmichael_statistical_1999}%
  \BibitemOpen
  \bibfield  {author} {\bibinfo {author} {\bibfnamefont {H.~J.}\ \bibnamefont
  {Carmichael}},\ }\href {//www.springer.com/de/book/9783540548829} {\emph
  {\bibinfo {title} {Statistical Methods in Quantum Optics 1: Master Equations
  and Fokker-Planck Equations}}},\ Theoretical and Mathematical Physics\
  (\bibinfo  {publisher} {Springer-Verlag},\ \bibinfo {address} {Berlin
  Heidelberg},\ \bibinfo {year} {1999})\BibitemShut {NoStop}%
\bibitem [{\citenamefont {Chang}\ and\ \citenamefont
  {Skinner}(1993)}]{chang_non-markovian_1993}%
  \BibitemOpen
  \bibfield  {author} {\bibinfo {author} {\bibfnamefont {T.~M.}\ \bibnamefont
  {Chang}}\ and\ \bibinfo {author} {\bibfnamefont {J.~L.}\ \bibnamefont
  {Skinner}},\ }\bibfield  {title} {\bibinfo {title} {Non-{Markovian}
  population and phase relaxation and absorption lineshape for a two-level
  system strongly coupled to a harmonic quantum bath},\ }\href
  {https://doi.org/10.1016/0378-4371(93)90489-Q} {\bibfield  {journal}
  {\bibinfo  {journal} {Physica A: Statistical Mechanics and its Applications}\
  }\textbf {\bibinfo {volume} {193}},\ \bibinfo {pages} {483} (\bibinfo {year}
  {1993})}\BibitemShut {NoStop}%
\bibitem [{\citenamefont {Ortíz}\ \emph {et~al.}(2019)\citenamefont {Ortíz},
  \citenamefont {Esmann},\ and\ \citenamefont
  {Lanzillotti-Kimura}}]{ortiz_phonon_2019}%
  \BibitemOpen
  \bibfield  {author} {\bibinfo {author} {\bibfnamefont {O.}~\bibnamefont
  {Ortíz}}, \bibinfo {author} {\bibfnamefont {M.}~\bibnamefont {Esmann}},\
  and\ \bibinfo {author} {\bibfnamefont {N.~D.}\ \bibnamefont
  {Lanzillotti-Kimura}},\ }\bibfield  {title} {\bibinfo {title} {Phonon
  engineering with superlattices: Generalized nanomechanical potentials},\
  }\href {https://doi.org/10.1103/PhysRevB.100.085430} {\bibfield  {journal}
  {\bibinfo  {journal} {Physical Review B}\ }\textbf {\bibinfo {volume}
  {100}},\ \bibinfo {pages} {085430} (\bibinfo {year} {2019})},\ \bibinfo
  {note} {publisher: American Physical Society}\BibitemShut {NoStop}%
\bibitem [{\citenamefont {Mahan}(2000)}]{mahan_many_2000}%
  \BibitemOpen
  \bibfield  {author} {\bibinfo {author} {\bibfnamefont {G.~D.}\ \bibnamefont
  {Mahan}},\ }\href@noop {} {\emph {\bibinfo {title} {Many Particle
  Physics}}},\ \bibinfo {edition} {3rd}\ ed.\ (\bibinfo  {publisher}
  {Springer},\ \bibinfo {year} {2000})\BibitemShut {NoStop}%
\end{thebibliography}


\newpage

\section{Appendix}

\renewcommand\thefigure{S\arabic{figure}}
\setcounter{figure}{0}
\renewcommand\theequation{S\arabic{equation}}
\setcounter{equation}{0}

\subsection{Numerical Details}\label{app:num_details}

For the discretization of phonon momentum, we use a Chebyshev-Gauss-type grid, which is dense around zero momentum and coarse for large momentum:
\begin{equation}
 \begin{split}
q_p = \Big(\frac{p}{2 N + 1} - \frac{1}{2 \pi} \textrm{sin}\Big(2 \pi \frac{p}{2 N + 1}\Big)\Big) 2 q_{\textrm{max}}
\end{split}
\end{equation}
with $p=1,...,N$. Note that while the grid approaches $q_p=0$, the point itself is not included to avoid numerical divergence. The exciton-phonon coupling efficiency for each mode $p$ is determined by the coupling factor $\alpha_p=\frac{M_p}{\hbar\omega_p}$ \cite{hohenester_quantum_2007,kaer_microscopic_2013}.
Following these references, we keep only one-phonon excitations with $\alpha_p > \varepsilon_{\textrm{cutoff}} \alpha_{\textrm{max}}$, two-phonon excitations with $\alpha_p \alpha_{p'} > \varepsilon_{\textrm{cutoff}} \alpha_{\textrm{max}}$ and three-phonon excitations with $\alpha_p \alpha_{p'} \alpha_{p''} > \varepsilon_{\textrm{cutoff}} \alpha_{\textrm{max}}$.
We use $\varepsilon_{\textrm{cutoff}}=0.01$, $q_{\textrm{max}}=6$ nm$^{-1}$, and $N=20$.
The number of 1-phonon, 2-phonon and 3-phonon states is $15$, $100$, and $284 $, respectively. The convergence of numerical results with respect to $N$ as well as the maximum total number of phonons $N_{\textrm{ph}}$ is discussed in the following.

\subsection{Convergence of polaron shift}\label{app:convergence}
\begin{figure}
\centering
\includegraphics[width=\columnwidth]{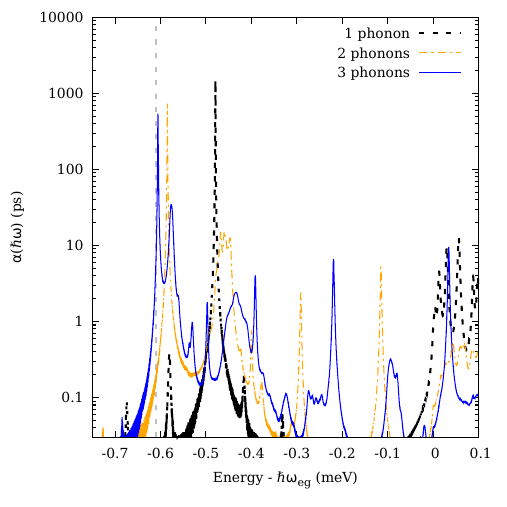}
\caption{Linear absorption spectrum of a WSe$_2$ QD coupled to 2D LA phonons at $T=0$ K for phonon lifetime $T_{\textrm{phon}}=300$ ps using phonon number states with different maximum total number of excitations. We compare cases where only 1 phonon can be excited, up to 2 phonons can be excited, and up to 3 phonons can be excited, respectively. Zero energy corresponds to the unperturbed QD transition $\hbar\omega_{\textrm{eg}}$. The vertical grey dashed line marks the analytic result for the polaron shift.}
\label{fig:spectrum_conv}
\end{figure}

To check the convergence of numerical results with respect to the maximum total number of phonons $N_{\textrm{ph}}$ in the computational basis states, we use the polaron shift as a figure of merit. Fig.~\ref{fig:spectrum_conv} shows linear optical spectra obtained by limiting the basis to many-particle states with $N_{\textrm{ph}}=1$ phonon, with up to $N_{\textrm{ph}}=2$ phonons and with up to $N_{\textrm{ph}}=3$ phonons, respectively. The emitter ZPL energy shifts to lower energies with increasing total number of phonons, as a larger maximum number of phonons allows for more (higher-order) emitter-phonon scattering channels.
Comparing the ZPL energies to the analytic result for the polaron shift Eq.~(\ref{eq:polaron_shift}), we find very good convergence for $N_{\textrm{ph}}=3$. Note that the PSB appears as a series of discrete peaks due to discretization of the phonon mode continuum.

\subsection{Convergence of Indistinguishability}\label{app:convergence_I}

\begin{figure}
\centering
\includegraphics[width=1.\columnwidth]{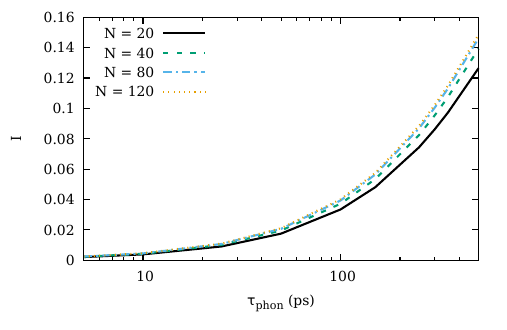}
\caption{Photon indistinguishability $\mathcal{I}$ of a WSe$_2$ QD coupled to 2D LA phonons at $T=0$ K in the absence of a cavity depending on the phonon lifetime $T_{\textrm{phon}}$ for different discretizations of the phonon mode continuum. $N$ is the number of discrete modes.}
\label{fig:I_g_0_conv}
\end{figure}

The quality of phonon discretization by means of the number $N$ of discrete modes is checked via the convergence of the indistinguishability in the absence of a cavity. As shown in Fig.~\ref{fig:I_g_0_conv}, we find good convergence already at $N=20$. Our interpretation is the following: While the PSB is poorly resolved in this case, it matters only at short time scales (see discussion in main text). In this regime, energy conservation is smeared out in a Fourier sense, which effectively broadens the discrete modes. At long time scales, where strict energy conservation is established, the pure dephasing rate dominates, which is already well converged at $N=20$.

\subsection{Photon indistinguishability from the augmented IBM}\label{app:IBM}

We start by deriving a solution to a generalization of the IBM \cite{mahan_many_2000}, which describes an emitter coupled to a phonon bath in the absence of a cavity (or, equivalently, in the limit of vanishing Jaynes-Cummings coupling). It is therefore defined by the Hamiltonian 
\begin{align}
    H_S &= H_{S,0} + V\,, \\
    \label{eq:H_S}
    H_{S,0} &= \hbar \omega_\mathrm{eg} \sigma_\mathrm{ee} + \sum_q \hbar\omega_q b_q^\dagger b_q^{\phantom\dagger} \,,
    \\
    V &=  \sigma_\mathrm{ee} \sum_q M_q(b_q^\dagger + b_q^{\phantom\dagger}) \equiv \sigma_\mathrm{ee} A\,,
\end{align}
which corresponds to the total Hamiltonian (\ref{eq:Hamiltonian_tot}) in the absence of photon degrees of freedom. The energy of the emitter ground state $\ket{\mathrm{g}}$ is chosen to be zero, while $\ket{\mathrm{e}}$ denotes the excited state. In the following, we also introduce the operators $\sigma^-=\ket{\textrm{g}}\bra{\textrm{e}}$ and $\sigma^+=\ket{\textrm{e}}\bra{\textrm{g}}$. 
As in our full density matrix calculation,
we take into account dissipative effects by including the interaction of the emitter-phonon system ($S$) with a reservoir ($R$). To this end, the IBM is augmented by 
Hamiltonians $H_{SR}$ describing the system-reservoir interaction and $H_R$ describing the reservoir itself. 
The total Hamiltonian is therefore given by
\begin{align}
    H=H_{S}+H_{SR}+H_R \,.
\label{eq:H_total}
\end{align}
The full quantum dynamics of the system-reservoir density matrix $\chi$ in the Schrödinger picture is formally given by
\begin{align}
    \chi(t) = e^{-\frac{i}{\hbar} H t} \chi(0) e^{\frac{i}{\hbar} H t} \,.
\label{eq:total_DM_dynamics}
\end{align}
However, in the presence of dissipation, the IBM can not be solved exactly but only approximately using the formalism of open quantum systems \cite{breuer_theory_2007}.
We are interested in the reduced system density matrix $\rho=\textrm{Tr}_R\left[ \chi \right]$, where $\textrm{Tr}_R\left[ ... \right] $ denotes the trace over reservoir degrees of freedom. Expressing the system-reservoir interaction in the form 
\begin{align}
    H_{SR} = \sum_{\nu} S_{\nu} B_{\nu}\,,
\label{eq:H_SR}
\end{align}
with system operators $ S_{\nu}$ and reservoir (or "bath") operators $B_{\nu} $, and treating it in Born-Markov approximaton, one can show that $\rho(t)$ obeys the von Neumann-Lindblad (vNL) equation \cite{lindblad_generators_1976, breuer_theory_2007}
\begin{equation}
 \begin{split}
\frac{\partial}{\partial t}\rho(t)= -\frac{i}{\hbar}\left[H_S, \rho(t) \right]+\sum_{\nu}\mathcal{L}_{\gamma_{\nu}}(S_{\nu})\rho(t)\,.
\end{split}
\label{eq:vNL_appendix}
\end{equation}
The dissipative Lindblad terms 
\begin{equation}
 \begin{split}
\mathcal{L}_{\gamma}(S)\rho=-\frac{\gamma}{2}\big[
S^{\dagger}S\rho+\rho S^{\dagger}S-2S\rho S^{\dagger}
\big] 
\end{split}
\label{eq:Lindblad_appendix}
\end{equation}
are defined by the system operators $ S_{\nu}$ and relaxation rates $\gamma_{\nu}$. In the following, we take into account the Lindblad terms $\mathcal{L}_{\Gamma}(\sigma^-)$ (radiative decay), $\mathcal{L}_{\gamma_q}(b_q)$ (phonon decay) and $\mathcal{L}_{\gamma_q^+}(b_q^\dagger)$ (phonon creation). Note that the phonon creation rate $\gamma_q^+$ is not included in the main text, since it vanishes in the limit of zero temperature. 

To obtain an expression for the photon indistinguishability $\mathcal{I}$, see Eq.~(\ref{eq:I}), we need to determine the two-time expectation value 
$C(t, \tau)=\braket{\sigma^+(t + \tau) \sigma^-(t)}$, which is formulated in the Heisenberg picture where
\begin{align}
    O(t) = e^{\frac{i}{\hbar} H t} O e^{-\frac{i}{\hbar} H t} \,.
\label{eq:heisenberg}
\end{align}
In the following, we make use of the linked cluster expansion method \cite{mahan_many_2000}, which requires to change to the interaction picture with respect to the emitter-phonon interaction $V$.
We denote operators in the interaction picture by a hat. Their dynamics is given by the "free" Hamiltonian $H_0$:
\begin{align}
    \hat{O}(t) &= e^{\frac{i}{\hbar} H_0 t} O e^{-\frac{i}{\hbar} H_0 t} \,, 
    \label{eq:interaction1} \\
    H_0&= H_{S,0}+H_{S,R}+H_R\,.
\label{eq:interaction2}
\end{align}
The above equation can not be solved exactly due to system-reservoir interaction. However, one can describe operator dynamics in Born-Markov approximation in terms of an adjoint vNL equation \cite{breuer_theory_2007}:
\begin{equation}
 \begin{split}
\frac{\partial}{\partial t}\hat{O}(t)= \frac{i}{\hbar}\left[H_{S,0}, O \right]+\sum_{\nu}\left[\mathcal{L}_{\gamma_{\nu}}(S_{\nu})\right]^{\dagger} O
\end{split}
\label{eq:operator_vNL}
\end{equation}
with
\begin{equation}
 \begin{split}
\left[\mathcal{L}_{\gamma}(S)\right]^{\dagger} O=-\frac{\gamma}{2}\big[
O S^{\dagger}S+ S^{\dagger}S O-2S^{\dagger}O S
\big] \,.
\end{split}
\label{eq:Lindblad_adjoint}
\end{equation}
We note that Eq.~(\ref{eq:operator_vNL}) strictly holds only for an operator under the expectation value. For basic operators, we obtain the solutions:
\begin{align}
\hat{\sigma}^{\pm}(t)&=\sigma^{\pm} e^{\pm i\omega_\mathrm{eg} t-\frac{\Gamma}{2} t}, 
\label{eq:basic_ops1}
\\
\hat{\sigma}_{\textrm{ee}}(t)&=\sigma_{\textrm{ee}} e^{-\Gamma t}, 
\label{eq:basic_ops2}
\\
\hat{b}_q^\dagger(t) &= b_q^\dagger e^{ i\omega_q t-\frac{\bar{\gamma}_q}{2} t},
\label{eq:basic_ops3}
\\
\hat{b}_q^{\phantom\dagger}(t)&=b_q^{\phantom\dagger}e^{ -i\omega_q t-\frac{\bar{\gamma}_q}{2} t}
\label{eq:basic_ops4}
\end{align}
with $\bar{\gamma}_q=\gamma_q - \gamma_q^+$. Operators are transformed from the interaction picture into the Heisenberg picture by means of the time evolution operator $U(t_2, t_1) = \mathcal{T} e^{-i \int_{t_1}^{t_2} dt^\prime \hat{V}(t^\prime)/\hbar}$:
\begin{align}
    O(t) = U(0,t) \hat{O}(t) U(t,0) \,,
\label{eq:int_to_heisenberg}
\end{align}
with $\mathcal{T}$ being the time-ordering operator. $U$ has the properties:
\begin{align}
U(t,0)&=e^{\frac{i}{\hbar} H_0 t}e^{-\frac{i}{\hbar} H t}, \\ U(0,t+\tau)&=U(0,t)U(t,t+\tau) \,.
\label{eq:U_prop}
\end{align}
This yields:
\begin{equation}
 \begin{split}
&C(t, \tau) \\
&= \braket{U(0,t + \tau) \hat{\sigma}^+(t + \tau) U(t + \tau, 0) U(0,t) \hat{\sigma}^-(t) U(t,0)} \\
&=
\mathrm{Tr}\left[
\chi(0)U(0,t)U(t,t+\tau)\hat{\sigma}^+(t + \tau)U(t + \tau, t)
\hat{\sigma}^-(t) U(t,0)
\right]\\
&=
\mathrm{Tr}\left[
U(t,0)\chi(0)U(0,t)
U(t,t+\tau)\hat{\sigma}^+(t + \tau)U(t + \tau, t)
\hat{\sigma}^-(t) 
\right]
.
\end{split}
\label{eq:C}
\end{equation}
From Eq.~(\ref{eq:basic_ops1}), we deduce that the time dependence of $\hat{\sigma}^-(t) $ is given by a simple exponential. Hence, in Eq.~(\ref{eq:C}) the operator $U(t + \tau, t)$ only acts on states $\propto\sigma^-\ket{...}$ where the emitter is in the ground state. 
On the other hand, $U$ contains the emitter-phonon coupling operator $\hat{V}\propto \sigma_{\textrm{ee}}$, which projects onto the excited emitter state, to arbitrary order. Thus only the zero-order term of $U$, which is the unity operator, remains:
\begin{equation}
 \begin{split}
&C(t, \tau) \\
&=
\mathrm{Tr}\left[
U(t,0)\chi(0)U(0,t)
U(t,t+\tau)\hat{\sigma}^+(t + \tau)\hat{\sigma}^-(t) 
\right] \\
&=
\mathrm{Tr}\left[
e^{\frac{i}{\hbar} H_0 t}e^{-\frac{i}{\hbar} H t}\chi(0)e^{\frac{i}{\hbar} H t}e^{-\frac{i}{\hbar} H_0 t}
U(t,t+\tau)\right.\\
&\left.\quad\quad\times\hat{\sigma}^+(t + \tau)\hat{\sigma}^-(t) 
\right]
.
\end{split}
\label{eq:C2}
\end{equation}
The initial density matrix $\chi(0)=\rho(0)R$ is given by a product of $\rho(0)$ and the (initial) reservoir density matrix $R$. According to the Born approximation, we assume that $\chi$ remains a product also at later times, with
\begin{equation}
 \begin{split}
&R(t)\equiv \frac{e^{-\beta H_R}}{\mathrm{Tr}\left[e^{-\beta H_R}\right]}
\end{split}
\label{eq:R}
\end{equation}
and the inverse temperature $\beta=(k_{\textrm{B}}T)^{-1}$.

Similar to the standard procedure given by Mahan \cite{mahan_many_2000}, we next apply a polaron transformation 
$\widetilde{O}=e^S O e^{-S}$
to separate emitter and phonon degrees of freedom in the expectation value. It turns out that due to the presence of dissipation, a modified transformation
\begin{equation}
 \begin{split}
S&=\sigma_{\textrm{ee}}\sum_q \frac{M_q}{\hbar}\left(
\frac{b_q^{\dagger}}{\omega_q-i\frac{\bar{\gamma_q}}{2}}-\frac{b_q^{\phantom\dagger}}{\omega_q+i\frac{\bar{\gamma_q}}{2}}
\right)\\
&\equiv \sigma_{\textrm{ee}}\sum_q \left(\lambda^\ast_q
b_q^{\dagger}-\lambda_q b_q^{\phantom\dagger}
\right)
\end{split}
\label{eq:S_mod}
\end{equation}
should be used. With the help of the identity
\begin{equation}
 \begin{split}
e^S O e^{-S}=O+\left[S,O\right]+\frac{1}{2!}\left[S,\left[S,O\right]\right]+...\,,
\end{split}
\label{eq:hadamard}
\end{equation}
one can derive the following operators in the polaron frame:
\begin{align}
e^S \sigma^{\pm} e^{-S}&=\sigma^{\pm} e^{\pm \sum_q \left(\lambda_q^\ast
b_q^{\dagger}-\lambda_q b_q^{\phantom\dagger}
\right) } \equiv \sigma^{\pm}X^{\pm}, 
\label{eq:polaron_ops1}
\\
e^S \sigma_{\textrm{ee}} e^{-S}&=\sigma_{\textrm{ee}}, 
\label{eq:polaron_ops2}
\\
e^S b_q^\dagger e^{-S} &= b_q^\dagger -\lambda_q \sigma_{\textrm{ee}},
\label{eq:polaron_ops3}
\\
e^S b_q^{\phantom\dagger} e^{-S} &= b_q^{\phantom\dagger} -\lambda^\ast_q \sigma_{\textrm{ee}}\,.
\label{eq:polaron_ops4}
\end{align}
Due to the dissipation-induced modification, the system Hamiltonian (\ref{eq:H_S}) obtains no simple form:
\begin{equation}
 \begin{split}
e^S H_S e^{-S}&=
H_{S,0}+\hbar\Delta_{\textrm{pol}}\sigma_{\textrm{ee}}\\
&+\sigma_{\textrm{ee}}\sum_q M_q\left( 
b_q^{\phantom\dagger}\frac{i\bar{\gamma}_q / 2}{\omega_q+i\bar{\gamma}_q / 2}-
b_q^{\dagger}\frac{i\bar{\gamma}_q / 2}{\omega_q-i\bar{\gamma}_q / 2}
\right)
\,.
\end{split}
\label{eq:polaron_H_S}
\end{equation}
On the other hand, a modified system Hamiltonian
\begin{equation}
 \begin{split}
\bar{H}_{S}=H_{S,0}+\sigma_{\textrm{ee}}\sum_q 
\hbar\omega_q (\lambda^\ast_q b_q^{\dagger}+\lambda_q b_q^{\phantom\dagger}) \equiv H_{S,0}+\sigma_{\textrm{ee}}\widetilde{A}
\end{split}
\label{eq:H_S_mod}
\end{equation}
becomes diagonal in the polaron frame:
\begin{equation}
 \begin{split}
e^S \bar{H}_S e^{-S}&=
H_{S,0}+\hbar\Delta_{\textrm{pol}}\sigma_{\textrm{ee}}
\,.
\end{split}
\label{eq:polaron_H_S_mod}
\end{equation}
Here, we have introduced the polaron shift, which is modified due to the dissipation of phonons:
\begin{equation}
 \begin{split}
\Delta_{\textrm{pol}}&=-\sum_{\bq} \frac{M_q^2}{\hbar^2} \frac{\omega_q}{\omega_q^2+(\bar{\gamma}_q/2)^2} \,.
\end{split}
\label{eq:polaron_shift_appendix}
\end{equation}
We now make some key assumptions for the initial system density matrix $\rho(0)$, such that it is particularly simple in the polaron frame.
First, we assume that the system density matrix initially is a product of emitter and phonon matrices:
$\rho(0)=\rho^0_{\textrm{em}}\rho^0_{\textrm{phon}}$. 
The emitter is assumed to be diagonal in $\lbrace \ket{\textrm{g}},\ket{\textrm{e}} \rbrace $, so that 
$e^S \rho^0_{\textrm{em}} e^{-S} =\rho^0_{\textrm{em}}$. The phonons are assumed to be thermal with respect to the modified interaction $\widetilde{A}$:
\begin{equation}
 \begin{split}
\rho^0_{\textrm{phon}}=\left(\mathrm{Tr}\left[ e^{-\beta (\sum_q \hbar\omega_q b_q^\dagger b_q^{\phantom\dagger} + \widetilde{A})} \right]\right)^{-1}e^{-\beta(\sum_q \hbar\omega_q b_q^\dagger b_q^{\phantom\dagger} + \widetilde{A})} .
\end{split}
\label{eq:rho_0_phon}
\end{equation}
Using Eq.~(\ref{eq:polaron_H_S_mod}) as well as the identity
\begin{equation}
 \begin{split}
e^S f(O) e^{-S}=f(e^S O e^{-S})
\end{split}
\label{eq:hadamard}
\end{equation}
for any function $f$, we find:
\begin{equation}
 \begin{split}
e^S\rho^0_{\textrm{phon}}e^{-S}&=\left(\mathrm{Tr}\left[ e^{-\beta (\sum_q \hbar\omega_q b_q^\dagger b_q^{\phantom\dagger} + \widetilde{A})}e^S e^{-S} \right]\right)^{-1}\\
&\times e^S e^{-\beta(\sum_q \hbar\omega_q b_q^\dagger b_q^{\phantom\dagger} + \widetilde{A})}e^{-S}\\
&=\left(\mathrm{Tr}\left[ e^{-\beta (\sum_q \hbar\omega_q b_q^\dagger b_q^{\phantom\dagger} + \Delta_{\textrm{pol}})}\right]\right)^{-1}\\
&\times e^{-\beta(\sum_q \hbar\omega_q b_q^\dagger b_q^{\phantom\dagger} + \Delta_{\textrm{pol}})}\\
&=\left(\mathrm{Tr}\left[ e^{-\beta \sum_q \hbar\omega_q b_q^\dagger b_q^{\phantom\dagger} }\right]\right)^{-1}e^{-\beta\sum_q \hbar\omega_q b_q^\dagger b_q^{\phantom\dagger}}\\
&=\rho^{\textrm{th}}_{\textrm{phon}}\,.
\end{split}
\label{eq:rho_0_phon_polaron}
\end{equation}
With the above assumptions, we can further evaluate the term $e^{-\frac{i}{\hbar} H t}\chi(0)e^{\frac{i}{\hbar} H t}$ in Eq.~(\ref{eq:C2}):
\begin{equation}
 \begin{split}
&\quad\,\, e^{-\frac{i}{\hbar} H t}\chi(0)e^{\frac{i}{\hbar} H t} \\
&= e^{-\frac{i}{\hbar} H t} R \rho^0_{\textrm{em}} \rho^0_{\textrm{phon}} e^{\frac{i}{\hbar} H t} \\
&=e^{-S}e^S e^{-\frac{i}{\hbar} H t} R e^{-S}e^S  \rho^0_{\textrm{em}} e^{-S}e^S\rho^0_{\textrm{phon}} e^{-S}e^S e^{\frac{i}{\hbar} H t} e^{-S}e^S \\
&=e^{-S}e^S e^{-\frac{i}{\hbar} H t} e^{-S} R \rho^0_{\textrm{em}} \rho^{\textrm{th}}_{\textrm{phon}}e^S e^{\frac{i}{\hbar} H t} e^{-S}e^S.
\end{split}
\label{eq:chi_t}
\end{equation}
The next step is to transform the total Hamiltonian (\ref{eq:H_total}):
\begin{equation}
 \begin{split}
   e^S H e^{-S} & = e^S H_{S} e^{-S} + H_R \\
   &+ e^S H_{SR,\textrm{em}} e^{-S} + e^S H_{SR,\textrm{phon}} e^{-S}  \,.
\end{split}
\label{eq:transf_H}
\end{equation}
The reservoir is not affected by the polaron transformation. We have split the system-reservoir interaction into two terms, which describe the effects on the emitter (radiative decay) and on the phonons (phonon decay and creation), respectively. 
According to Eq.~(\ref{eq:H_SR}), these terms transform as
\begin{equation}
 \begin{split}
   e^S H_{SR} e^{-S} & = \sum_{\nu} e^S S_{\nu} e^{-S} B_{\nu}  \,.
\end{split}
\label{eq:transf_H_SR}
\end{equation}
In the following, we assume that the effect of the polaron transformation on the emitter-reservoir coupling is negligible. Then the action of $e^S H e^{-S}$ on the system density matrix would be described in Born-Markov approximation by the vNL equation:
\begin{equation}
 \begin{split}
\frac{\partial}{\partial t}\rho(t)& = -\frac{i}{\hbar}\left[e^S H_S e^{-S}, \rho(t) \right]+
\mathcal{L}_{\Gamma}(\sigma^-)\rho(t) \\
+&\sum_{q}\mathcal{L}_{\gamma_q}(e^S b_q^{\phantom\dagger} e^{-S})\rho(t)
+\sum_{q}\mathcal{L}_{\gamma^+_q}(e^S b_q^{\dagger} e^{-S})\rho(t)\\
& = -\frac{i}{\hbar}\left[H_{S,0}+\Delta_{\textrm{pol}}\sigma_{\textrm{ee}}, \rho(t) \right]+
\mathcal{L}_{\Gamma}(\sigma^-)\rho(t) \\
+&\sum_{q}\mathcal{L}_{\gamma_q}(b_q^{\phantom\dagger})\rho(t)
+\sum_{q}\mathcal{L}_{\gamma^+_q}(b_q^{\dagger})\rho(t)
\,.
\end{split}
\label{eq:vNL_mod}
\end{equation}
In the second line we have made use of Eqs.~(\ref{eq:polaron_H_S}), (\ref{eq:polaron_ops3}), and (\ref{eq:polaron_ops4}). We can therefore approximately replace $e^S  H e^{-S}$ in Eq.~(\ref{eq:chi_t}) by $H_{S,0}+\Delta_{\textrm{pol}}\sigma_{\textrm{ee}}+H_{SR}+H_R=H_0+\Delta_{\textrm{pol}}\sigma_{\textrm{ee}}  $. Note that $H_0$ contains the system-reservoir interaction, see Eq.~(\ref{eq:interaction2}). By inserting Eq.~(\ref{eq:chi_t}) into Eq.~(\ref{eq:C2}), we obtain:
\begin{equation}
 \begin{split}
C(t, \tau)
&=
\mathrm{Tr}\left[
e^{\frac{i}{\hbar} H_0 t}
e^{-S} e^{-\frac{i}{\hbar} (H_0+\Delta_{\textrm{pol}}\sigma_{\textrm{ee}}) t}
R \rho^0_{\textrm{em}} \rho^{\textrm{th}}_{\textrm{phon}}\right.\\
&\times 
e^{\frac{i}{\hbar} (H_0+\Delta_{\textrm{pol}}\sigma_{\textrm{ee}}) t}
e^S
e^{-\frac{i}{\hbar} H_0 t}
U(t,t+\tau)\\
&\left.\times\hat{\sigma}^+(t + \tau)\hat{\sigma}^-(t) 
\right]\\
&=
\mathrm{Tr}\left[
e^{\frac{i}{\hbar} H_0 t}
e^{-S} e^{-\frac{i}{\hbar} H_0 t}
R_\textrm{em} \rho^0_{\textrm{em}}\right. \\
&\times 
e^{\frac{i}{\hbar} H_0 t}
\rho^{\textrm{th}}_{\textrm{phon}}
R_\textrm{phon}
e^S
e^{-\frac{i}{\hbar} H_0 t}
U(t,t+\tau)\\
&\left.\times\hat{\sigma}^+(t + \tau)\hat{\sigma}^-(t) 
\right]
.
\end{split}
\label{eq:C3}
\end{equation}
Here, we have assumed that the thermal phonon density matrix $\rho^{\textrm{th}}_{\textrm{phon}} $ commutes with $H_{SR}$ and $H_R$. Moreover, we have factorized the reservoir density matrix $R$ into a part $R_\textrm{em}$ interacting with the emitter and a phonon part $R_\textrm{phon}$.

We expand the (adjoint) time-evolution operator in Eq.~(\ref{eq:C3}) into an exponential series:
\begin{equation}
 \begin{split}
U(t, t+\tau) &= U^{\dagger}(t+\tau,t)=  \bar{\mathcal{T}} e^{i \int_{t}^{t+\tau} dt^\prime \hat{V}^{\dagger}(t^\prime)/\hbar} \\
&=\sum_{n=0}^\infty \Big(\frac{i}{\hbar}\Big)^n\frac{1}{n!} \int_t^{t + \tau} dt_1 \dots \int_t^{t + \tau} dt_n \\
&\times \bar{\mathcal{T}} \left[ \hat{\sigma}_{\textrm{ee}}(t_1)\hat{A}(t_1)...\hat{\sigma}_{\textrm{ee}}(t_n)\hat{A}(t_n) \right]
\end{split}
\label{eq:U_adjoint}
\end{equation}
with $\bar{\mathcal{T}}$ being the anti-time-ordering operator. As a consequence, Eq.~(\ref{eq:C3}) may be written as
\begin{equation}
    C(t, \tau) = \sum_{n=0}^\infty C_n(t, \tau).
    \label{eq:C_series}
\end{equation}
In the $n$-th order term, we have the expression
\begin{equation}
 \begin{split}
 &e^{-\frac{i}{\hbar} H_0 t}
\bar{\mathcal{T}} \left[ \hat{\sigma}_{\textrm{ee}}(t_1)\hat{A}(t_1)...\hat{\sigma}_{\textrm{ee}}(t_n)\hat{A}(t_n) \right]
\hat{\sigma}^+(t + \tau)\hat{\sigma}^-(t)
e^{\frac{i}{\hbar} H_0 t} 
\\
=&\bar{\mathcal{T}} \left[
e^{-\frac{i}{\hbar} H_0 t} \hat{\sigma}_{\textrm{ee}}(t_1)e^{\frac{i}{\hbar} H_0 t}e^{-\frac{i}{\hbar} H_0 t} \hat{A}(t_1)e^{\frac{i}{\hbar} H_0 t}e^{-\frac{i}{\hbar} H_0 t}... \right.
\\
 \times&\hat{\sigma}_{\textrm{ee}}(t_n)e^{-\frac{i}{\hbar} H_0 t}e^{\frac{i}{\hbar} H_0 t}\hat{A}(t_n)e^{\frac{i}{\hbar} H_0 t}e^{-\frac{i}{\hbar} H_0 t} \Big ] \\
\times&\hat{\sigma}^+(t + \tau)e^{\frac{i}{\hbar} H_0 t}e^{-\frac{i}{\hbar} H_0 t}\hat{\sigma}^-(t)e^{\frac{i}{\hbar} H_0 t} \\
=&
\bar{\mathcal{T}}\left[ \hat{\sigma}_{\textrm{ee}}(t_1-t)\hat{A}(t_1-t)...\hat{\sigma}_{\textrm{ee}}(t_n-t)\hat{A}(t_n-t) \right]
\hat{\sigma}^+(\tau)\sigma^-\,, 
\end{split}
\label{eq:U_adjoint_H_0_t}
\end{equation}
where we have used Eq.~(\ref{eq:interaction1}). The anti-time-ordering operator only sorts operators at times $t_i$ but does not affect the operators at times $t$ and $t+\tau$. By shifting the time integration variables $t_1\rightarrow t_1 + t,...,t_n\rightarrow t_n + t$, we obtain:
\begin{equation}
 \begin{split}
C_n(t, \tau)
&=
\mathrm{Tr}\left[
e^{-S} e^{-\frac{i}{\hbar} H_0 t}
R_\textrm{em} \rho^0_{\textrm{em}}e^{\frac{i}{\hbar} H_0 t}\right. \\
&\times 
\rho^{\textrm{th}}_{\textrm{phon}}
R_\textrm{phon}
e^S
\Big(\frac{i}{\hbar}\Big)^n\frac{1}{n!} \int_0^{\tau} dt_1 \dots \int_0^{\tau} dt_n \\
&\left.\times \bar{\mathcal{T}} \left[ \hat{\sigma}_{\textrm{ee}}(t_1)\hat{A}(t_1)...\hat{\sigma}_{\textrm{ee}}(t_n)\hat{A}(t_n) \right]\hat{\sigma}^+(\tau)\sigma^- 
\right]
\,,
\end{split}
\label{eq:C_n}
\end{equation}
where the time evolution in $t$ and $\tau$ is disentangled. To enable a separation of emitter and phonon degrees of freedom, we neglect the time dependence of the $\hat{\sigma}_{\textrm{ee}}$ operators, which is caused by radiative decay according to Eq.~(\ref{eq:basic_ops2}). As a consequence, Eq.~(\ref{eq:C_n}) contains a product of $n$ operators $\sigma_{\textrm{ee}}$, which acts on a state $\propto\sigma^+\ket{...}$ as a unity operator. Similarly, we can analyze the action of the operator $e^{-S} e^{-\frac{i}{\hbar} H_0 t}
R_\textrm{em} \rho^0_{\textrm{em}}e^{\frac{i}{\hbar} H_0 t}
\rho^{\textrm{th}}_{\textrm{phon}}
R_\textrm{phon}
e^S $ on a state $\propto\sigma^+\ket{...}$: 
Since $\rho^0_{\textrm{em}}$ has been assumed to be diagonal, the density matrix $e^{-\frac{i}{\hbar} H_0 t}
R_\textrm{em} \rho^0_{\textrm{em}}e^{\frac{i}{\hbar} H_0 t} $ remains diagonal and therefore commutes with $e^{-S}$, which is proportional to $\sigma_{\textrm{ee}}$ (see Eq.~(\ref{eq:S_mod})). We can thus replace the operators $S$ by $\widetilde{S}=\sum_q \left(\lambda_q^\ast
b_q^{\dagger}-\lambda_q b_q^{\phantom\dagger}
\right) $.
Due to this replacement, the coefficients $C_n$ factorize into emitter and phonon parts, yielding
\begin{equation}
 \begin{split}
    &C_n(t,\tau) = C^{\textrm{em}}(t,\tau)C^{\textrm{phon}}_n(\tau) \\ =& 
    \mathrm{Tr}_{\textrm{em}}\left[
e^{-\frac{i}{\hbar} H_{\textrm{em}} t}
R_\textrm{em} \rho^0_{\textrm{em}}e^{\frac{i}{\hbar} H_{\textrm{em}} t} \hat{\sigma}^+(\tau)\sigma^-  \right] \\
\times &
\mathrm{Tr}_{\textrm{phon}}\left[
\rho^{\textrm{th}}_{\textrm{phon}}
R_\textrm{phon}
e^{\widetilde{S}}
\Big(\frac{i}{\hbar}\Big)^n\frac{1}{n!} \int_0^{\tau} dt_1 \dots \int_0^{\tau} dt_n \right.\\
&\left.\quad\quad\times \bar{\mathcal{T}} \hat{A}(t_1)...\hat{A}(t_n) e^{-\widetilde{S}}
\right]
\end{split}
\label{eq:C_n_fact}
\end{equation}
with $H_{\textrm{em}}= \hbar \omega_\mathrm{eg} \sigma_\mathrm{ee} + H_{SR,\textrm{em}}+H_{R,\textrm{em}}$ and $\hat{\sigma}^+(\tau)=e^{i/\hbar H_{\textrm{em}} \tau} \sigma^+ e^{-i/\hbar H_{\textrm{em}} \tau} $.

The solution for the emitter part
\begin{equation}
 \begin{split}
    C^{\textrm{em}}(t,\tau) &=
    \mathrm{Tr}_{\textrm{em}}\left[
e^{-\frac{i}{\hbar} H_{\textrm{em}} t}
R_\textrm{em} \rho^0_{\textrm{em}}e^{\frac{i}{\hbar} H_{\textrm{em}} t} \hat{\sigma}^+(\tau)\sigma^-  \right] \\
&=
    \mathrm{Tr}_{\textrm{em}}\left[
R_\textrm{em} \rho^0_{\textrm{em}}\hat{\sigma}^+(t+\tau)\hat{\sigma}^-(t)\right] \\
&=
    \braket{\hat{\sigma}^+(t+\tau)  \hat{\sigma}^-(t)}_{\textrm{em}}
    \\
    &= \braket{\sigma_\mathrm{ee}}(0)e^{i\omega_\mathrm{eg} \tau - \Gamma (t + \tau/2)}
\end{split}
\label{eq:C_em_final}
\end{equation}
follows from a corresponding vNL equation with the help of the quantum regression formula (see main text). Here, the system Hamiltonian $\hbar \omega_\mathrm{eg} \sigma_\mathrm{ee}$ is augmented by the Lindblad term $\mathcal{L}_{\Gamma}(\sigma^-)$.

The phonon part of $C_n(t,\tau)$ is further evaluated by applying the (modified) polaron transformation:
\begin{equation}
 \begin{split}
    &C^{\textrm{phon}}_n(\tau) \\ =& 
\mathrm{Tr}_{\textrm{phon}}\left[
\rho^{\textrm{th}}_{\textrm{phon}}
R_\textrm{phon}
\Big(\frac{i}{\hbar}\Big)^n\frac{1}{n!} \int_0^{\tau} dt_1 \dots \int_0^{\tau} dt_n \right.\\
&\left.\quad\quad\times \bar{\mathcal{T}} e^{\widetilde{S}}\hat{A}(t_1)...\hat{A}(t_n) e^{-\widetilde{S}}
\right]\\ =& 
\mathrm{Tr}_{\textrm{phon}}\left[
\rho^{\textrm{th}}_{\textrm{phon}}
R_\textrm{phon}
\Big(\frac{i}{\hbar}\Big)^n\frac{1}{n!} \int_0^{\tau} dt_1 \dots \int_0^{\tau} dt_n \right.\\
&\left.\quad\quad\times \bar{\mathcal{T}} e^{\widetilde{S}}\hat{A}(t_1)e^{-\widetilde{S}}e^{\widetilde{S}}...e^{-\widetilde{S}}e^{\widetilde{S}}\hat{A}(t_n) e^{-\widetilde{S}}
\right]\,.
\end{split}
\label{eq:C_n_phon}
\end{equation}
We transform $\hat{A}(t)$ by making use of the operator time dependence including dissipation, see Eqs.~(\ref{eq:basic_ops3}) and (\ref{eq:basic_ops4}):
\begin{equation}
 \begin{split}
    &e^{\widetilde{S}}\hat{A}(t) e^{-\widetilde{S}}\\
    =& \sum_q M_q e^{\widetilde{S}}\left(b_q^\dagger e^{i\omega_q t} + b_q^{\phantom\dagger} e^{-i\omega_q t} \right)e^{-\widetilde{S}} e^{-\frac{\bar{\gamma}_q}{2}t}\\
    =& \sum_q M_q \left((b_q^\dagger-\lambda_q) e^{i\omega_q t} + (b_q^{\phantom\dagger}-\lambda^\ast_q) e^{-i\omega_q t} \right) e^{-\frac{\bar{\gamma}_q}{2}t}
    \\
    =& \hat{A}(t) - \sum_q\frac{M_q}{\hbar}\left(  
    \frac{e^{i\omega_q t}}{\omega_q+i\frac{\bar{\gamma}_q}{2}} + \frac{e^{-i\omega_q t}}{\omega_q-i\frac{\bar{\gamma}_q}{2}}
    \right)e^{-\frac{\bar{\gamma}_q}{2}t}
    \\
    \equiv& \hat{A}(t) - f(t)\,.
\end{split}
\label{eq:A_transform}
\end{equation}
Denoting $\mathrm{Tr}_{\textrm{phon}}\left[
\rho^{\textrm{th}}_{\textrm{phon}}
R_\textrm{phon}...\right]=\braket{...}_\mathrm{phon}$, we obtain
\begin{equation}
 \begin{split}
    &C^{\textrm{phon}}_n(\tau) =
   \Big(\frac{i}{\hbar}\Big)^n\frac{1}{n!} \int_0^{\tau} dt_1 \dots \int_0^{\tau} dt_n
\\
&\times \braket{\bar{\mathcal{T}} (\hat{A}(t_1)-f(t_1))...(\hat{A}(t_n)-f(t_n))}_\mathrm{phon}
\,.
\end{split}
\label{eq:C_n_phon2}
\end{equation}
At this point, the actual \textit{linked cluster expansion} starts. Since $\braket{...}_\mathrm{phon}$ is a thermal expectation value with respect to a bilinear Hamiltonian, we can apply Wick's theorem to $C^{\textrm{phon}}_n(\tau) $
 \cite{mahan_many_2000}, expanding the $n$-operator expectation values into all possible products of pair-wise expectation values:
\begin{equation}
 \begin{split}
    &\braket{\bar{\mathcal{T}} (\hat{A}(t_1)-f(t_1))...(\hat{A}(t_n)-f(t_n))}_\mathrm{ph} \\ = \sum_{\mathrm{all\,products}} &\braket{\bar{\mathcal{T}} (\hat{A}(t_i)-f(t_i))(\hat{A}(t_j)-f(t_j))}_\mathrm{ph}\\
    \times&\dots \\
    \times&\braket{\bar{\mathcal{T}}(\hat{A}(t_k)-f(t_k))(\hat{A}(t_l)-f(t_l))}_\mathrm{ph}.
\end{split}
\label{eq:Wick}
\end{equation}
As expectation values with unequal numbers of phonon creation and annihilation operators vanish, we have to distinguish terms with even and odd $n$.
For odd $n=2m+1$, the coefficients have the following form:
\begin{equation}
 \begin{split}
    &C^{\textrm{phon}}_{2m+1}(\tau) =
   \Big(\frac{i}{\hbar}\Big)^{2m+1}\frac{1}{(2m+1)!} \Bigg[ \\ &\sum_{\mathrm{all\,prod.}} \left(\iint_0^{\tau} dt_idt_j \braket{\bar{\mathcal{T}} \hat{A}(t_i)\hat{A}(t_j)}_\mathrm{ph}\right)^m \int_0^{\tau} dt_k (-f(t_k))
\\
+&\sum_{\mathrm{all\,prod.}} \left(\iint_0^{\tau} dt_idt_j  \braket{\bar{\mathcal{T}} \hat{A}(t_i)\hat{A}(t_j)}_\mathrm{ph}\right)^{m-1} \left(\int_0^{\tau} dt_k (-f(t_k))\right)^3\\
+&...+\left(\int_0^{\tau} dt_k (-f(t_k))\right)^{2m+1}\Bigg] \\
&=\Big(\frac{i}{\hbar}\Big)^{2m}\frac{1}{m!}\big(\phi(\tau)\big)^m\big(-i F(\tau)\big) \\
&+\Big(\frac{i}{\hbar}\Big)^{2m-2}\frac{1}{(m-1)!}\big(\phi(\tau)\big)^{m-1}\big(-i F(\tau)\big)^3\frac{1}{3!}\\
&+...+\big(-i F(\tau)\big)^{2m+1}\frac{1}{(2m+1)!}
\,.
\end{split}
\label{eq:Wick_odd}
\end{equation}
First of all, we have used the fact that the pairwise expectation values as well as the functions $f$ only differ in their respective time integration variables. In the first class of factorizations, there are $2m+1$ possibilities to choose the index $t_k$. The remaining $2m$ $\hat{A}$-operators can be grouped in $(2m-1)\cdot(2m-3)\cdot ...\cdot 1=(2m)!/(2^m m!)$ different combinations. In the second class, there are $\binom{2m+1}{3}=(2m+1)!/((2m-2)!3!)$ ways to pick $3$ out of the $2m+1$ functions $f$ and $(2m-3)\cdot(2m-5)\cdot ...\cdot 1=(2m-2)!/(2^{m-1} (m-1)!)$ combinations for the $\hat{A}$-operators, and so on. Finally, there is only one term containing all $f$. We have also introduced the shorthand expressions
\begin{align}
F(\tau)&=\frac{1}{\hbar}\int_0^\tau dt f(t), \label{eq:def_F} \\
\phi(\tau)&= \frac{1}{2} \iint_0^{\tau} dt_idt_j \braket{\bar{\mathcal{T}} \hat{A}(t_i)\hat{A}(t_j)}_\mathrm{ph}\,.
\label{eq:def_phi}
\end{align}
For even $n=2m$, a similar rationale yields
\begin{equation}
 \begin{split}
    &C^{\textrm{phon}}_{2m}(\tau)
=\Big(\frac{i}{\hbar}\Big)^{2m}\frac{1}{m!}\big(\phi(\tau)\big)^m \\
&+\Big(\frac{i}{\hbar}\Big)^{2m-2}\frac{1}{(m-1)!}\big(\phi(\tau)\big)^{m-1}\big(-i F(\tau)\big)^2\frac{1}{2!}\\
&+...+\big(-i F(\tau)\big)^{2m}\frac{1}{(2m)!}
\,.
\end{split}
\label{eq:Wick_even}
\end{equation}
Summation over all $C^{\textrm{phon}}_{n}(\tau)$ leads to:
\begin{equation}
 \begin{split}
    &\sum_{n=0}^{\infty} C^{\textrm{phon}}_{n}(\tau)
=1-iF(\tau)+\Big(\frac{i}{\hbar}\Big)^{2}\phi(\tau)+\frac{1}{2!}\big(-iF(\tau)\big)^{2} \\
&+\Big(\frac{i}{\hbar}\Big)^{2}\phi(\tau)\big(-iF(\tau)\big)+\frac{1}{3!}\big(-iF(\tau)\big)^{3}+\Big(\frac{i}{\hbar}\Big)^{4}\frac{1}{2!}\big(\phi(\tau)\big)^{2}\\
&+\frac{1}{4!}\big(-iF(\tau)\big)^{4}+\Big(\frac{i}{\hbar}\Big)^{2}\phi(\tau)\frac{1}{2!}\big(-iF(\tau)\big)^{2}+...\\
&=\sum_{n=0}^{\infty} \frac{1}{n!}\big(-iF(\tau)\big)^{n} \sum_{n'=0}^{\infty} \Big(\frac{i}{\hbar}\Big)^{n'}\frac{1}{n'!}\big(\phi(\tau)\big)^{n'}\\
&=e^{-iF(\tau)}e^{-\frac{1}{\hbar^2}\phi(\tau)}
\,.
\end{split}
\label{eq:C_series_explicit}
\end{equation}
Due to the separation of emitter and phonon degrees of freedom, the linked cluster expansion assumes a particularly simple form, with $iF(\tau)+\frac{1}{\hbar^2}\phi(\tau)$ being the only distinct linked cluster.

We proceed by evaluating the function $\phi(\tau)$, which corresponds to a double time integral over a phonon Green function:
\begin{equation}
 \begin{split}
\phi(\tau)&= \frac{1}{2} \iint_0^{\tau} dt_idt_j \braket{\bar{\mathcal{T}} \hat{A}(t_i)\hat{A}(t_j)}_\mathrm{ph}\\
&=\frac{1}{2}\sum_q M_q^2 \iint_0^{\tau} dt_idt_j \\
&\times\Big[ \braket{\bar{\mathcal{T}} \hat{b}^{\dagger}_q(t_i)\hat{b}^{\phantom\dagger}_q(t_j)}_\mathrm{ph}
+\braket{\bar{\mathcal{T}} \hat{b}^{\phantom\dagger}_q(t_i)\hat{b}^{\dagger}_q(t_j)}_\mathrm{ph}\Big]
\,.
\end{split}
\label{eq:eval_phi1}
\end{equation}
Here, $\bar{\mathcal{T}}$ moves the operator with later time to the right.
Choosing $t_j=t_i+\tau>t_i$, the two-time expectation values can be evaluated via the vNL equation for a system Hamiltonian $\sum_q \hbar\omega_q b^{\dagger}_q b^{\phantom\dagger}_q $ augmented by Lindblad terms $\mathcal{L}_{\gamma_q}(b_q)$ and $\mathcal{L}_{\gamma_q^+}(b_q^\dagger)$:
\begin{equation}
 \begin{split}
\braket{ \hat{b}^{\dagger}_q(t_i)\hat{b}^{\phantom\dagger}_q(t_j)}_\mathrm{ph}=&
\braket{ \hat{b}^{\dagger}_q(t_i)\hat{b}^{\phantom\dagger}_q(t_i+\tau)}_\mathrm{ph}\\
=&N_q e^{-i\omega_q\tau}e^{-\frac{\bar{\gamma}_q}{2}\tau}\,, \\
\braket{ \hat{b}^{\phantom\dagger}_q(t_i)\hat{b}^{\dagger}_q(t_j)}_\mathrm{ph}=&
(N_q+1) e^{i\omega_q\tau}e^{-\frac{\bar{\gamma}_q}{2}\tau}
\,,
\end{split}
\label{eq:QRT_phonon}
\end{equation}
where we have again used the quantum regression formula. The appearance of the Bose function
$ N_q =1/(e^{\beta \hbar \omega_q} - 1)$
is a consequence of the phonons and the external reservoir being thermal, which implies the Boltzmann-type relation $\gamma_q^+ = e^{-\beta \hbar \omega_q}\gamma_q$ between phonon decay and creation rates.
Similar results are obtained for $t_i=t_j+\tau>t_j$. We thus obtain
\begin{equation}
 \begin{split}
\phi(\tau)&=\frac{1}{2}\sum_q M_q^2 \iint_0^{\tau} dt_idt_j \\
&\times\Big[ 
N_q e^{(-i\omega_q-\frac{\bar{\gamma}_q}{2})|t_i-t_j|}
+(N_q+1) e^{(i\omega_q-\frac{\bar{\gamma}_q}{2})|t_i-t_j|}
\Big] \\
&=\sum_q M_q^2\left[
\frac{N_q}{i\omega_q+\frac{\bar{\gamma}_q}{2}}-\frac{1+N_q}{i\omega_q-\frac{\bar{\gamma}_q}{2}}
\right]\tau \\
&+\sum_q M_q^2 \left[
\frac{N_q}{\big(i\omega_q+\frac{\bar{\gamma}_q}{2}\big)^2}\big(e^{(-i\omega_q-\frac{\bar{\gamma}_q}{2})\tau}-1\big)\right. \\
&+\left.
\frac{1+N_q}{\big(i\omega_q-\frac{\bar{\gamma}_q}{2}\big)^2}\big(e^{(i\omega_q-\frac{\bar{\gamma}_q}{2})\tau}-1\big)
\right]
\,.
\end{split}
\label{eq:eval_phi2}
\end{equation}
Using this together with the result for $F(\tau)$,
\begin{equation}
 \begin{split}
F(\tau)&=\int_0^\tau dt 
\sum_q\frac{M_q}{\hbar^2}\left(  
    \frac{e^{i\omega_q t}}{\omega_q+i\frac{\bar{\gamma}_q}{2}} + \frac{e^{-i\omega_q t}}{\omega_q-i\frac{\bar{\gamma}_q}{2}}
    \right)e^{-\frac{\bar{\gamma}_q}{2}t} \\
    &= 
    \sum_q \frac{M_q^2}{\hbar^2} \left[
\frac{1}{i\big(\omega_q+i\frac{\bar{\gamma}_q}{2}\big)^2}\big(e^{(i\omega_q-\frac{\bar{\gamma}_q}{2})\tau}-1\big)\right. \\
&-\left.
\frac{1}{i\big(\omega_q-i\frac{\bar{\gamma}_q}{2}\big)^2}\big(e^{(-i\omega_q-\frac{\bar{\gamma}_q}{2})\tau}-1\big)
\right]    
    \,,
\end{split}
\label{eq:eval_F}
\end{equation}
we obtain the final result for the phonon part of $C(t,\tau)$ from Eq.~(\ref{eq:C_series_explicit}):
\begin{equation}
 \begin{split}
    &\sum_{n=0}^{\infty} C^{\textrm{phon}}_{n}(\tau)
=e^{i\Delta_{\textrm{pol}}\tau-\frac{\gamma_{\textrm{pure}}}{2}\tau+\Phi(\tau)}
\,.
\end{split}
\label{eq:C_series_final}
\end{equation}
Here we have used the polaron shift (\ref{eq:polaron_shift_appendix}) and introduced 
the pure dephasing rate 
\begin{align}
    \frac{\gamma_\mathrm{pure}}{2} = \sum_q \frac{M_q^2}{\hbar^2} \frac{(2N_q + 1)\bar{\gamma}_q/2}{\omega_q^2 + (\bar{\gamma}_q/2)^2}
    \label{eq:gamma_pure_full}
\end{align}
as well as the phonon dephasing integral
\begin{equation}
 \begin{split}
    \Phi(\tau) = \sum_q \frac{M_q^2}{\hbar^2} \Big[ &\frac{N_q + 1}{(\omega_q - i\bar{\gamma}_q/2)^2} (e^{(-i\omega_q - \bar{\gamma}_q/2)\tau}-1) 
    \\
    + & \frac{N_q}{(\omega_q + i\bar{\gamma}_q/2)^2} (e^{(i\omega_q - \bar{\gamma}_q/2)\tau}-1) \Big]\,.
\end{split}
\label{eq:Phi_tau_full}
\end{equation}
Compared to the standard IBM, both the polaron shift and the phonon dephasing integral obtain corrections due to phonon decay and creation. Even more, a completely new pure dephasing term emerges, which is proportional to $\bar{\gamma}_q$. These results are in agreement with the findings of Zimmermann and Runge \cite{zimmermann_dephasing_2002}, with the interesting detail that \textit{half} the Lindblad rate $\bar{\gamma}_q = \gamma_q-\gamma^+_q$ enters as characteristic decay rate. At zero temperature, the Bose function in Eqs.~(\ref{eq:gamma_pure_full}) and (\ref{eq:Phi_tau_full}) vanishes and $\bar{\gamma}_q$ is given by the phonon decay rate $\gamma_q$.

Combining Eqs.~(\ref{eq:C_em_final}) and (\ref{eq:C_series_final}), the final result for the two-time expectation value is 
\begin{align}
    C(t,\tau) = \braket{\sigma_\mathrm{ee}} (0) e^{ i(\omega_\mathrm{eg} + \Delta_\mathrm{pol})\tau  -\Gamma (t+\tau/2) - \frac{\gamma_\mathrm{pure}}{2} \tau + \Phi(\tau)} \,.
\end{align}
We can use this expression to evaluate the indistinguishability formula (\ref{eq:I}):
\begin{equation}
 \begin{split}
\mathcal{I}_{g=0}&=
\frac{\int_0^{\infty}\,dt\int_0^{\infty}\,d\tau\,|\big< \sigma^+(t+\tau)\sigma^-(t) \big>|^2}{\int_0^{\infty}\,dt\int_0^{\infty}\,d\tau\,\big< \sigma^+(t+\tau)\sigma^-(t+\tau) \big>\big< \sigma^+(t)\sigma^-(t) \big>}\Big|_{g=0}
\\
&=\frac{\int_0^{\infty}\,dt\int_0^{\infty}\,d\tau\,|C(t,\tau)|^2}{\int_0^{\infty}\,dt\int_0^{\infty}\,d\tau\,C(t+\tau,0)C(t,0)}
\\
&=\frac{\int_0^{\infty}\,dt\int_0^{\infty}\,d\tau\,
e^{-\Gamma(2t+\tau)-\gamma_{\textrm{pure}}\tau+2\textrm{Re}\,\Phi(\tau)}
}{\int_0^{\infty}\,dt\int_0^{\infty}\,d\tau\,e^{-\Gamma(2t+\tau)}}
\\
&=\Gamma\int_0^{\infty} d\tau e^{-(\Gamma\tau+\gamma_{\textrm{pure}}\tau-2\textrm{Re}\,\Phi(\tau))}\,.
\end{split}
\label{eq:I_IBM_appendix}
\end{equation}

\end{document}